\documentclass[lettersize,journal]{IEEEtran}
\usepackage{amsmath,latexsym,amssymb,stmaryrd,pifont}
\usepackage{setspace,stackrel,tikz,graphicx}
\usepackage{exscale,relsize,subfig,textcomp,stackrel,setspace,float}
\usepackage{mdframed,pifont}
\usepackage{multicol,xfrac}
\usepackage{multirow, booktabs,placeins}
\usepackage{mathtools}
\usepackage{tabularx}
\usepackage[input-decimal-markers=.]{siunitx}
\usepackage{longtable}
\usepackage[T1]{fontenc}
\usepackage{caption}
\usepackage{fancyhdr}
\usepackage{lipsum}
\usepackage{multicol}
\usepackage{graphicx}
\usepackage[utf8]{inputenc}
\pdfoutput=1
\usepackage{amsthm}
\usepackage{array}
\usepackage{colortbl}
\usepackage{xcolor}
\usepackage{mathtools,amssymb,lipsum}
\usepackage{subfig}

\usepackage{cuted}
\usepackage{soul}
\usepackage[nolist]{acronym}

\begin{acronym}
    \acro{MIMO}{multiple-input, multiple-output}
    \acro{AWGN}{additive white Gaussian noise}
    \acro{KPI}{key performance indicator}
    \acro{RCS}{radar cross section}
    \acro{ISAC}{integrated sensing and communications}
    \acro{Tx}{transmitter}
    \acro{Rx}{receiver}
    \acro{PL}{path loss}
    \acro{PDF}{probability density function}
    \acro{CDF}{cumulative distribution function}
    \acro{GoF}{goodness-of-fit}
    \acro{3GPP}{3rd Generation Partnership Project}
    \acro{USRP}{universal software radio peripheral}
    \acro{InF}{indoor factory}
    \acro{InH}{indoor hotspot}
    \acro{UAV}{unmanned aerial vehicle}
    \acro{AMR}{autonomous mobile robot}
    \acro{RA}{robotic arm}
    \acro{LoS}{line-of-sight}
    \acro{H}{horizontal}
    \acro{KS}{Kolmogorov-Smirnov}
    \acro{NF}{near-field}
    \acro{MFE}{mean fitting error}
    \acro{MSE}{mean squared error}
    \acro{ZC}{Zadoff-Chu}
    \acro{IF}{intermediate frequency}
    \acro{QR}{quadruped robot}
    \acro{CIR}{channel impulse response}
    \acro{THz}{terahertz}
    \acro{FI}{floating intercept}
    \acro{RFIC}{radio-frequency integrated circuit}
    \acro{RF}{radio-frequency}
    \acro{NYU}{New York University}
    \acro{ADC}{analog-to-digital converter}
\end{acronym}

\begin{document}

\title{\vspace{0cm}\LARGE Statistical and Deterministic RCS Characterization for ISAC Channel Modeling\vspace{0em}}
\makeatletter
\patchcmd{\@maketitle}
  {\addvspace{0\baselineskip}\egroup}
  {\addvspace{0\baselineskip}\egroup}
  {}
  {}
\makeatother

\author{Ali Waqar Azim, Ahmad Bazzi, Roberto Bomfin, Nikolaos Giakoumidis, Theodore S. Rappaport, Marwa Chafii
\thanks{This work is supported in part by the NYUAD Center for Artificial Intelligence and Robotics, funded by Tamkeen under the Research Institute Award CG010.\\
Ali Waqar Azim, Ahmad Bazzi, Roberto Bomfin, and Marwa Chafii are with the Engineering Division, New York University Abu Dhabi (NYUAD), 129188, UAE
(email: {ali.waqar.azim,ahmad.bazzi,roberto.bomfin, marwa.chafii}@nyu.edu). Ahmad Bazzi, Theodore S. Rappaport and Marwa Chafii are with NYU WIRELESS, NYU Tandon School of Engineering, Brooklyn, 11201, NY, USA (email: tsr@nyu.edu). Nikolaos Giakoumidis is  KINESIS Lab, Core Technology Platforms, NYUAD, 129188, UAE and with Intelligent Systems Lab, Cultural Technology and Communication, University of the Aegean, 811 00 Mitilini, Greece (email: giakoumidis@nyu.edu).}}
\maketitle
\begin{abstract}
In this study, we perform a statistical analysis of the radar cross section (RCS) for various test targets in an indoor factory at \(25\)-\(28\) GHz, with the goal of formulating parameters that may be used for target identification and other sensing applications for future wireless systems. 
\textcolor{black}{The analysis is conducted based on measurements in monostatic and bistatic configurations for bistatic angles of} \(20^\circ\), \(40^\circ\), and \(60^\circ\), \textcolor{black}{which are functions of transmitter-receiver (T-R) and target positions, via accurate} \(3\)dB beamwidth of \(10^\circ\) \textcolor{black}{in both azimuth and elevation planes.}
The test targets include unmanned aerial vehicles, an autonomous mobile robot, and a robotic arm.
We utilize parametric statistical distributions to fit the measured RCS data. The analysis reveals that the \textcolor{black}{\textit{lognormal and gamma distributions}} are effective in modeling the RCS of the test targets \textcolor{black}{over different reflecting points of the target itself, i.e. when target is in motion.}
 Additionally, we provide a framework for evaluating the deterministic bistatic RCS of a rectangular sheet \textcolor{black}{of laminated wood, due to its widespread use in indoor hotspot environments.}
\textcolor{black}{Novel deterministic and statistical} RCS models are evaluated, incorporating dependencies on the bistatic angle, T-R distance (\(2\)m -\(10\)m) and the target. The results demonstrate that some proposed RCS models accurately fit the measured data, highlighting their applicability in bistatic configurations.
\end{abstract}
\IEEEpeerreviewmaketitle
\vspace{-4mm}
\section{Introduction}
\Ac{ISAC} combines wireless communications and radar sensing to efficiently utilize scarce radio resources and shared hardware for dual purposes, with two primary configurations: monostatic and bistatic \cite{9449071}. 
\begin{figure}[t]
    \centering
    \includegraphics[width=1\linewidth]{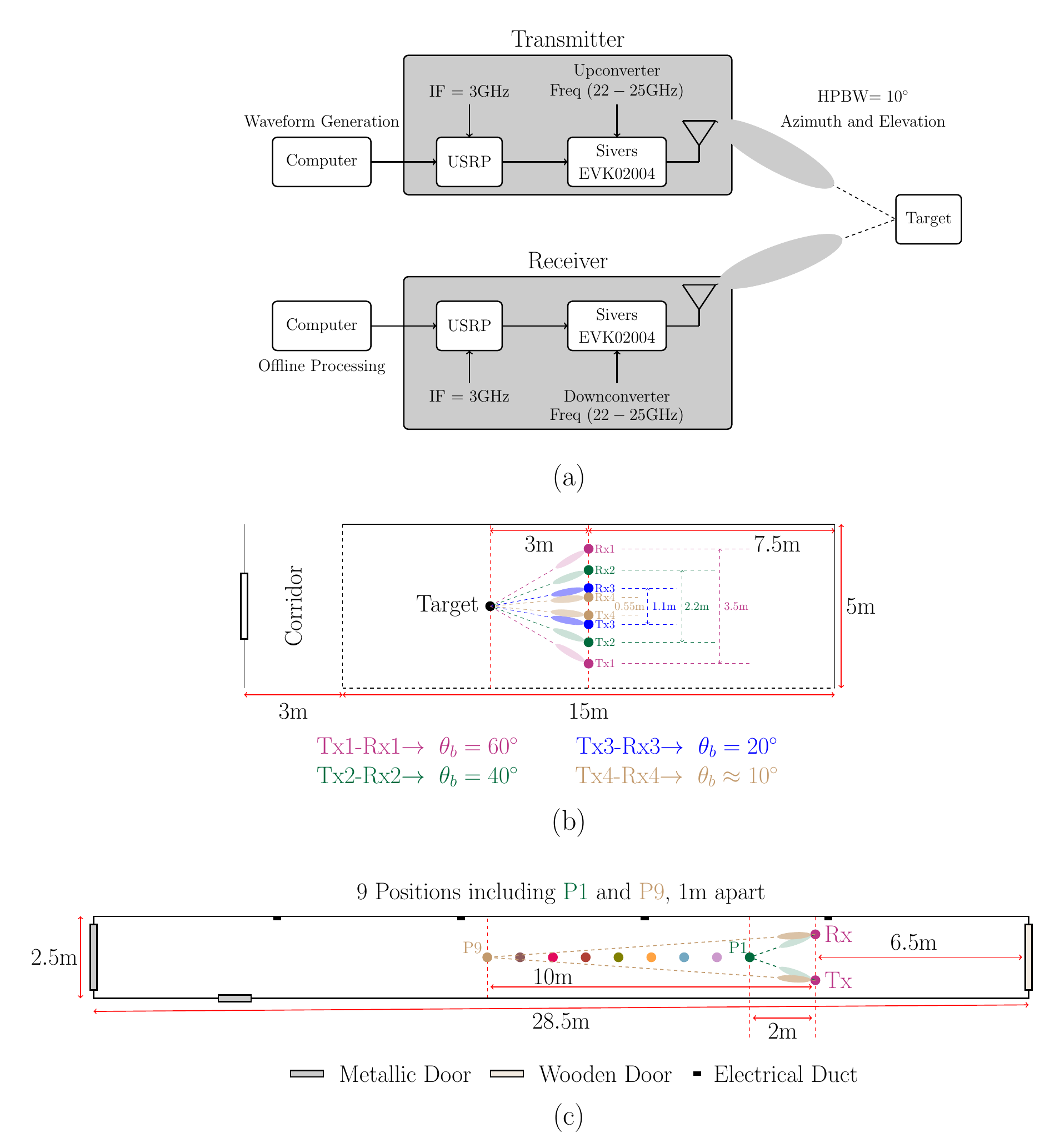} 
    \caption{\textcolor{black}{(a) Testbed for the RCS measurement, (b) the   layout of the measurement environment for statistical RCS analysis, (c)  the   layout of the measurement environment for deterministic RCS analysis.  }}
    \label{fig1} 
\end{figure}
For sensing, the target's \ac{RCS} characterization is an important aspect of ISAC channel modeling. The \ac{RCS} is an intrinsic property of the target and is independent of the radar system, determined by factors such as the geometric structure, the material composition, and the operating frequency, among others \cite{wei2024integrated, levanon1988radar, cheng1989field}.
The \ac{RCS} can be modeled using deterministic or statistical approaches \cite{wei2024integrated}. Deterministic {\ac{RCS}} modeling utilizes numerical and geometric techniques to compute the {\ac{RCS}} with precision. In addition, deterministic {\ac{RCS}} modeling considers various elements, such as the structural characteristics of the test object and the operating frequency. In contrast, statistical {\ac{RCS}} modeling represents the RCS as a random variable characterized by a specific {\ac{PDF}} \cite{borden1983statistical}. 

    \subsection{\textcolor{black}{Measurement Setup \& Considerations}}
\textcolor{black}{Our measurement testbed is described in Fig. {\ref{fig1}(a)}, which provides a block diagram representation of the bistatic system model used for {\ac{RCS}} measurements. Our testbed contains a signal generation, upconversion, transmission, and a symmetrical reception process. The system employs a \ac{USRP} for waveform generation and digitization, with the Sivers EVK02004 handling {\ac{RF}} upconversion and downconversion. The measurement operates over a frequency range of {$25$}-{$28$} GHz, with an {\ac{IF}} of {$3$} GHz. The \ac{Tx} and \ac{Rx} antennas are aligned to ensure proper signal reflection and capture. The half-power beamwidth is \(10^\circ\) in elevation and azimuth. The \ac{Tx} and \ac{Rx} radio-frequency integrated circuits (RFICs) are interfaced with \(\mathrm{B205mini}\) \acp{USRP}, having a bandwidth of \(20\) MHz, each. The \ac{Tx} employs a known \ac{ZC} sequence of length \(128\) as the transmitted signal, generated using MATLAB. At the \ac{Rx}, the signal is processed to extract the \ac{CIR}. In order to change the output frequency of the RFIC, the upconverter and downconverter frequencies are changed in steps of $1$GHz. Subsequently, the \acp{RFIC} at both \ac{Tx} and \ac{Rx} are calibrated for the updated frequencies.}

Figs. {\ref{fig1}(b)} and {\ref{fig1}(c)} provide a top-view representation of the measurement setups, including antenna orientations, distances, and bistatic angles. Fig. {\ref{fig1}(b)} illustrates the statistical RCS characterization setup, showing four distinct Tx-Rx pairs positioned at different bistatic angles with the boresight of the antenna tilted towards the target. Fig. {\ref{fig1}(c)} depicts the deterministic RCS setup in a corridor, where the target moves along predefined positions (P1 to P9), altering the bistatic angle. In both setups, the \ac{Tx} and \ac{Rx} are carefully aligned to ensure accurate signal reflection and capture over the specified distances.
Our \ac{InF} environment is the KINESIS Lab, Core Technology Platforms, located two floors below the Arts Center at the \ac{NYU} Abu Dhabi campus, in Saadiyat Island, Abu Dhabi, the United Arab Emirates. 
The \ac{InF} layout is \(5\) m in width, \(15\) m in length and \(8.5\) m in height, which is sufficient to fly small and medium sized \acp{UAV}.
The \ac{InH} environment is an indoor corridor nearby the KINESIS lab, which is \(2.5\) m in width, \(28.5\) m in length and \(2.5\) m in height.
\textcolor{black}{The total amount of data collected throughout our entire campaign is \(3.544\) gigabytes.}
\textcolor{black}{For the statistical case, \textit{the \ac{RCS} is measured during motion of the target, which can capture a statistical representation of the target in question.}}

Our statistical \ac{RCS} analysis includes two {\acp{UAV}} that differ in size and material composition, \textcolor{black}{and are observed in distances ranging from} \(3\)m to \(3.5\)m from the {\ac{Tx}}/{\ac{Rx}}.
For robotic systems, we examine an {\ac{AMR}} and a {\ac{RA}}.
For {\acp{UAV}}, the \ac{RCS} is measured during flight and rotational motion, while for the {\ac{AMR}} and {\ac{RA}}, the data is collected as they emulate tasks typically performed in an {\ac{InF}} setting. All measurements were conducted in {\ac{LoS}} conditions with {\ac{H}} polarization for both the {\ac{Tx}} and {\ac{Rx}}. Parametric distributions are employed to model the measured \ac{RCS} data for statistical analysis\textcolor{black}{, via {\ac{GoF}} metrics, which are the {\ac{KS}} statistic and the {\ac{MSE}}, which provide quantitative measures of the accuracy of the fit.} 
\textcolor{black}{For sensitivity, the \ac{ADC} of the \ac{USRP} has a resolution of $12$ bits, while the \ac{ADC} of the \ac{RFIC} has a resolution of $10$ bits.}

\textcolor{black}{In addition to statistical modeling of \ac{RCS} for common factory objects, we also sought to determine the \ac{NF} \ac{RCS} of common objects found in factories, such as a rectangular sheet of laminated wood in an {\ac{InH}} environment for operating frequencies ranging from $25$ GHz to $28$ GHz, via a deterministic \ac{RCS} framework.}
\textcolor{black}{Laminated wood is widely used in modern offices for everything from desktops and meeting tables to cabinets and wall cladding, thanks to its durability, range of aesthetic finishes, and easy maintenance.}
 {\Ac{PL}} measurements are collected across various bistatic angles to evaluate the \ac{RCS} deterministically. We propose multiple \ac{RCS} models that depend on the bistatic angle and the distance between the {\ac{Tx}}/{\ac{Rx}} and the target’s center. 
 \textcolor{black}{Our deterministic model is a modified version of the {\ac{FI}} model, in a sense that it accounts twice for double {\ac{PL}} and introduces an {\ac{RCS}} term (see Section {\ref{sec3b}} for more details). After fitting onto the modified {\ac{FI}} model, we assess different {\ac{NF}} {\ac{RCS}} expressions to find the best {\ac{NF}} {\ac{RCS}} in terms of the lowest {\ac{MFE}}.}
 All measurements are conducted under {\ac{LoS}} conditions, with {\ac{H}}-polarization for {\ac{Tx}} and {\ac{Rx}}. In addition to fitting the {\ac{RCS}}, other parameters such as the {\ac{PL}} exponent, intercept, and shadowing factor are also estimated. The fitting results indicate a strong dependence of the {\ac{RCS}} on the bistatic angle.
 
\subsection{\textcolor{black}{Standardization Progress and \ac{RCS}}}
\textcolor{black}{While {\ac{RCS}} was considered decades ago in the early days of the cellphone industry for channel modeling, the sparsity of large buildings in macrocells made the approach much less valuable than today's applications, where the excellent temporal resolution (in the order of nanoseconds) and antenna directionality offered by the small wavelengths of millimeter wave and terahertz frequencies makes the use of {\ac{RCS}} viable for {\ac{ISAC}} applications}  \cite{108383,8761205}.
 \textcolor{black}{Using the {\ac{RCS}} method to identify objects is} particularly effective when the backscatter echo power from the target exhibits fluctuations due to variations in its scattering properties or changes in its cross-sectional areas \cite{8761205}. The measured \ac{RCS} data \textcolor{black}{may then be} fitted to parametric statistical distributions, \textcolor{black}{for lookup tables and future use in learning and optimization as more measurements are collected for the identified object(s).}
 The distribution(s) demonstrating robust {\ac{GoF}} metrics with the measured data are selected to represent the target’s {\ac{RCS}}. Importantly, the {\ac{3GPP}} endorses statistical \ac{RCS} modeling for \ac{ISAC} channel modeling \cite{3gpp_r1_2408100}, highlighting its significance in capturing dynamic and fluctuating target behaviors, besides channel modeling for the upper mid-band as part of $6$G \cite{Shakya2025}.

%
\vspace{-2mm}
\section{Statistical RCS Characterization of Indoor Factory Targets}\label{sec2}
 In this section, we provide the statistical {\ac{RCS}} characterization of {\ac{InF}} targets. 
%

\subsection{Indoor \textcolor{black}{Factory} Targets}\label{sec2a}
\begin{figure}[t]
    \centering
    \subfloat[Mavic \(2\) Pro]{
        \includegraphics[width=0.45\columnwidth]{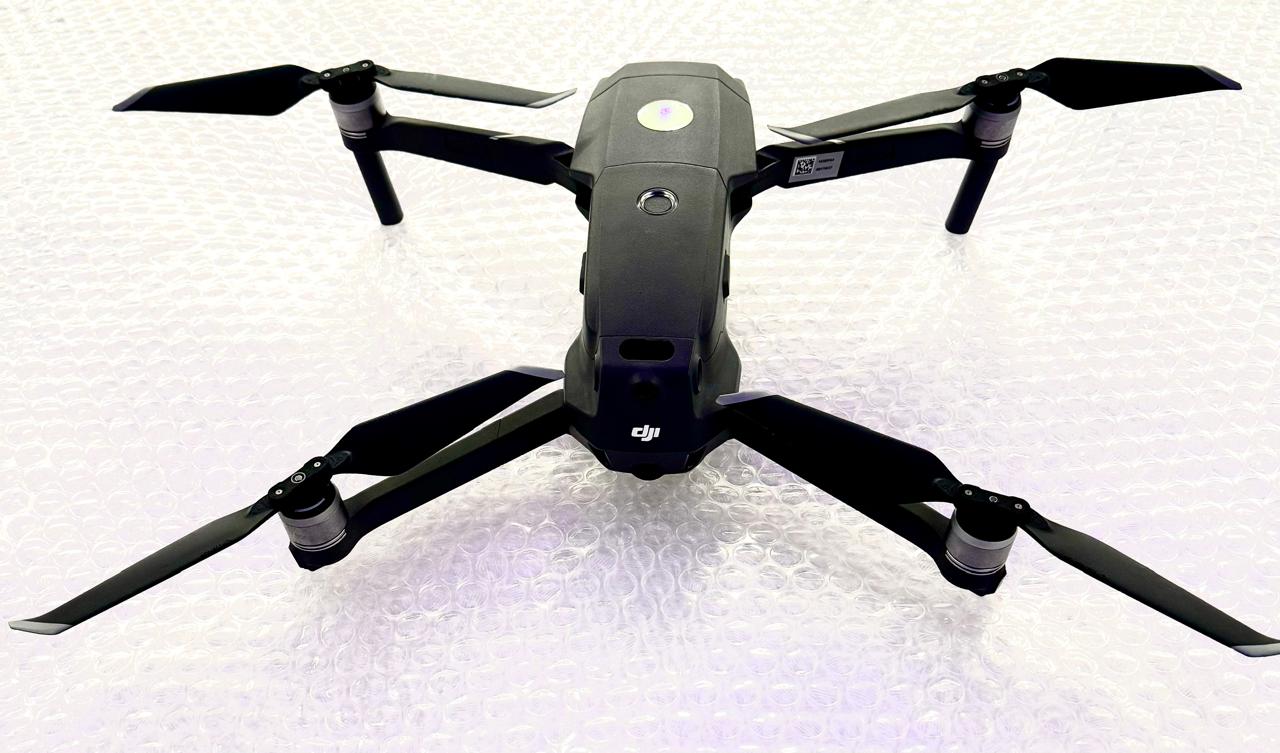}
        \label{drone01}
    }
    \hfill
    \subfloat[Matrice \(300\) RTK]{
        \includegraphics[width=0.43\columnwidth, height=2.4cm, keepaspectratio]{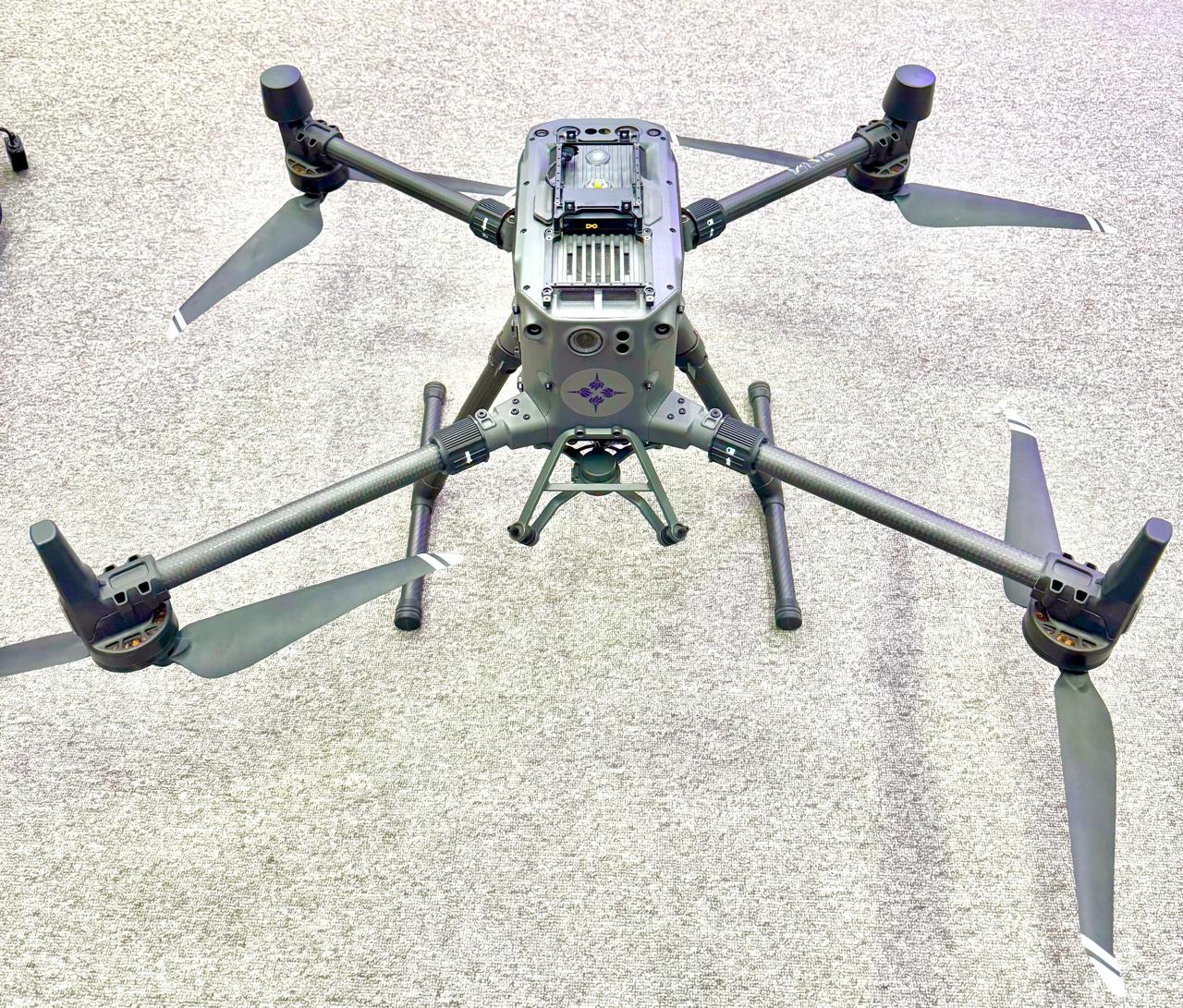}
        \label{drone02}
    }
    \caption{{\ac{UAV}} test targets used in the study.}
    \label{drones}
\end{figure}

The \ac{InF} environment, shown in Fig. {\ref{fig1}(b)}, is the KINESIS Lab, Core Technology Platforms, which is $2$ floors below the Arts Center at \ac{NYU} Abu Dhabi campus. The \ac{InF} layout is \(5\) m in width, \(15\) m in length and \(8.5\) m in height.
In our measurements, the targets we consider are commonly found in {\ac{InF}} environments, \textcolor{black}{which include} (i) two types of {\acp{UAV}}; (ii) an {\ac{RA}}; and (iii) an {\ac{AMR}}, referred to as the {\ac{QR}}. The specifications of the targets are given in Table \ref{equipment_table}.
\begin{table*}[ht]
\small 
\centering
\caption{Specifications of Equipment Used in Measurements}
\begin{tabular}{|p{3.7cm}|p{5cm}|p{1.7cm}|p{5cm}|}
\hline
\textbf{Equipment} & \textbf{Dimensions} & \textbf{Mass} & \textbf{Materials} \\ \hline
DJI Mavic \(2\) Pro & Folded: \(214\times 91 \times 84\)mm & \(907\)g & Magnesium alloy, reinforced plastic, \\
&Unfolded: \(322 \times 242 \times 84\) mm&&carbon fiber, glass, silicon components\\\hline
DJI Matrice \(300\) RTK & Folded: \(430\times 420 \times 430\) mm & \(3.6\)-\(6.3\)kg & Carbon fiber-reinforced plastic, \\ 
& Unfolded: \(810 \times 670 \times 430\) mm&&aluminum, plastic \\\hline
Robotic Arm & Reach: \(820\) mm & \(\approx 30\)kg & Aluminum, titanium, steel, plastic, \\ 
(KUKA LBR iiwa 14 R\(820\)) &Height: \(1640\) mm (with arm)&& polymer composites\\\hline
Quadruped Robot & Length: \(1100\) mm; Width: \(500\) mm; & \(\approx 32.7\)kg & Aluminum, titanium, carbon fiber  \\ 
(Boston Dynamics Spot)& Height: \(610\) mm (walking without arm) &&composites, polymer\\\hline
\end{tabular}
\label{equipment_table}
\end{table*}
The unfolded configurations of the {\acp{UAV}} are presented in Fig. \ref{drones}(a) and (b). Fig. \ref{drones}(a) illustrates the Mavic \(2\) Pro, while Fig. \ref{drones}(b) displays the Matrice \(300\) RTK. \textcolor{black}{The Mavic} \(2\) \textcolor{black}{Pro and Matrice} \(300\) \textcolor{black}{RTK {\acp{UAV}}'} respective folded and unfolded dimensions are detailed in Table \ref{equipment_table}. In addition to the {\ac{UAV}} size, different orientations of the {\acp{UAV}} play a critical role in characterizing the {\ac{RCS}} dependency. Therefore, measurements were conducted with the {\acp{UAV}} in flight, rotating steadily above the observation point, \textcolor{black}{which ensures capturing backscatter reflections at {\ac{Rx}} from all possible {\ac{UAV}} orientations}. The vertical distance between the ground and the bottom of the {\acp{UAV}} is also different due to their dimensional differences. For Mavic \(2\) Pro, this distance was approximately \(0.9\)m, while for Matrice \(300\) RTK, it was approximately \(0.6\)m. Another factor influencing the {\ac{RCS}} is the placement of the lithium-ion battery. For the Matrice \(300\) RTK, the battery is mounted on one side, whereas for the Mavic \(2\) Pro, the battery is positioned on top of the {\ac{UAV}}. \textcolor{black}{The design difference of both drones} is expected to result in stronger specular reflections from the Matrice \(300\) RTK, compared to the Mavic \(2\) Pro. \textcolor{black}{Note that} {\acp{UAV}} were flying over the observation point during the measurement process.
The \textit{Observation point} is located \(3\)m away from the center of the line segment connecting {\ac{Tx}} and {\ac{Rx}}. The {\ac{Tx}} and {\ac{Rx}} bore-sight beams also converge at that observation point.
%
\begin{figure}[t]
    \centering
    \subfloat[Robotic Arm]{
        \includegraphics[width=0.6\columnwidth, height=4.6cm, keepaspectratio]{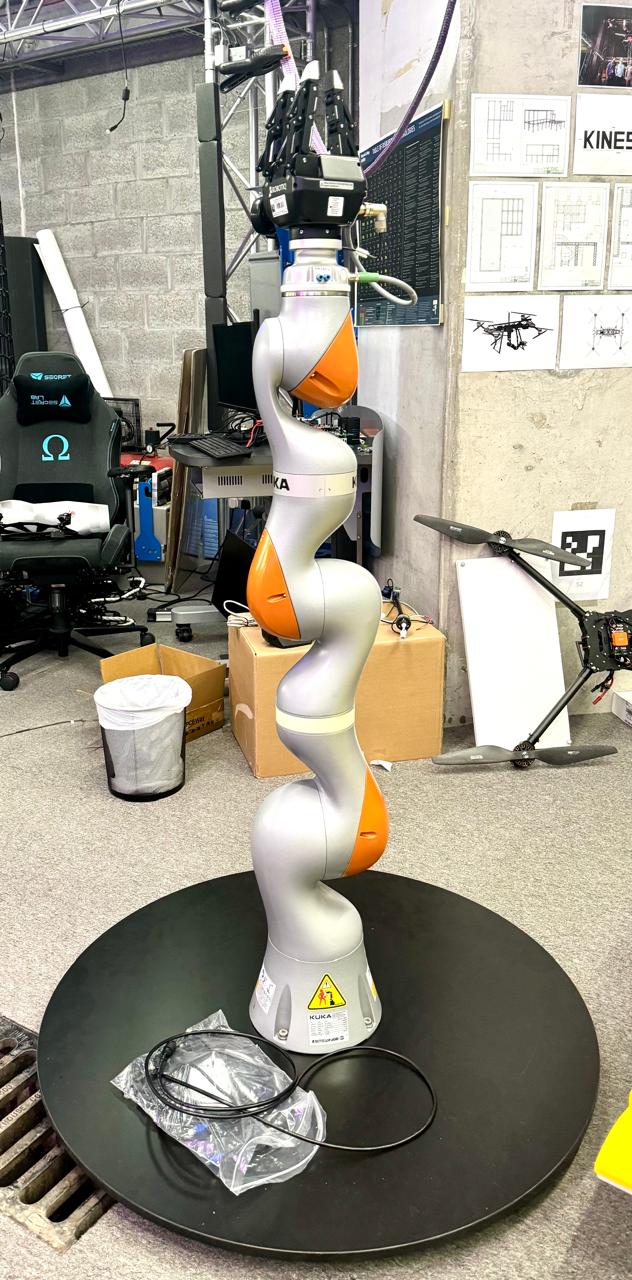}
        \label{robotic_arm}
    }
    \hfill
    \subfloat[Quadruped Robot]{
        \includegraphics[width=0.43\columnwidth]{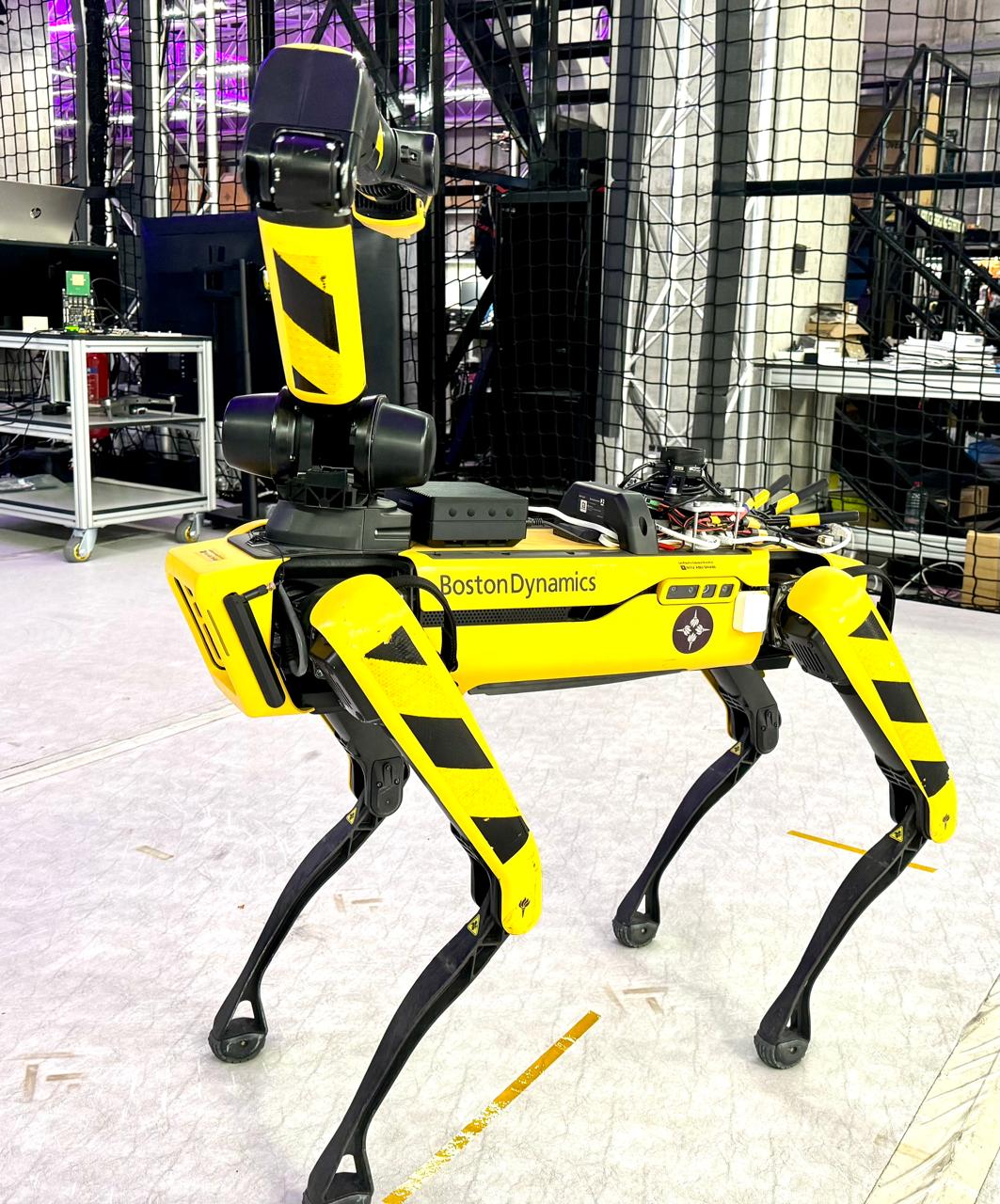}
        \label{robot}
    }
    \caption{InF equipment used in the study: (a) {\ac{RA}}, (b) {\ac{QR}}.}
    \label{targets}
\end{figure}
%


The second target under consideration is the {\ac{RA}}, as shown in Fig. \ref{robotic_arm}. The reach and height specifications of the {\ac{RA}} are detailed in Table \ref{equipment_table}. 
\textcolor{black}{Measurements were conducted with {\ac{RA}} positioned at the observation point and continuously moving, utilizing its seven joints to perform various tasks characteristic of its operation in an {\ac{InF}} setting, including pick-and-place operations, etc.} 
The {\ac{RA}}'s maximum achievable height is \(1640\) mm. However, it can be significantly reduced during specific motions, such as those that simulate pick-and-place tasks, thereby decreasing the reflective surface area. 

The third target is the {\ac{QR}}, depicted in Fig. \ref{robot}. The dimensions of the {\ac{QR}} are provided in Table \ref{equipment_table}. {\ac{RCS}} measurements were conducted while the {\ac{QR}} was in motion around the observation point, with two distinct types of movements analyzed: (i) lateral motion along the observation point and (ii) longitudinal motion towards and away from the observation point. \textcolor{black}{The lateral and longitudinal motions} replicate the typical operational patterns of {\acp{QR}} in {\ac{InF}}. The lateral motion captures the robot's side profile, while the longitudinal motion focuses on the front/back profile, providing {\ac{RCS}} measurements across different orientations. A non-conducting object of \(20\) cm height was placed at the observation point. The {\ac{QR}} moved vertically up and down the non-conducting object.
\vspace{-2mm}
\subsection{Experimental Setup and {\ac{RCS}} Evaluation Methodology}\label{sec2b}
\begin{figure}[t]
    \centering
    \includegraphics[width=0.3\textwidth]{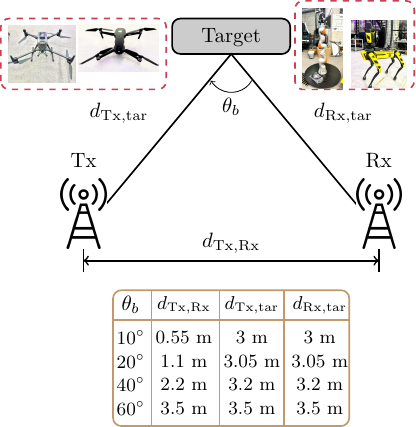} 
    \caption{An illustration of the measurement setup configuration.}
    \label{measurement_setup} 
\end{figure}

\begin{figure}[t]
    \centering
    \includegraphics[width=0.3\textwidth]{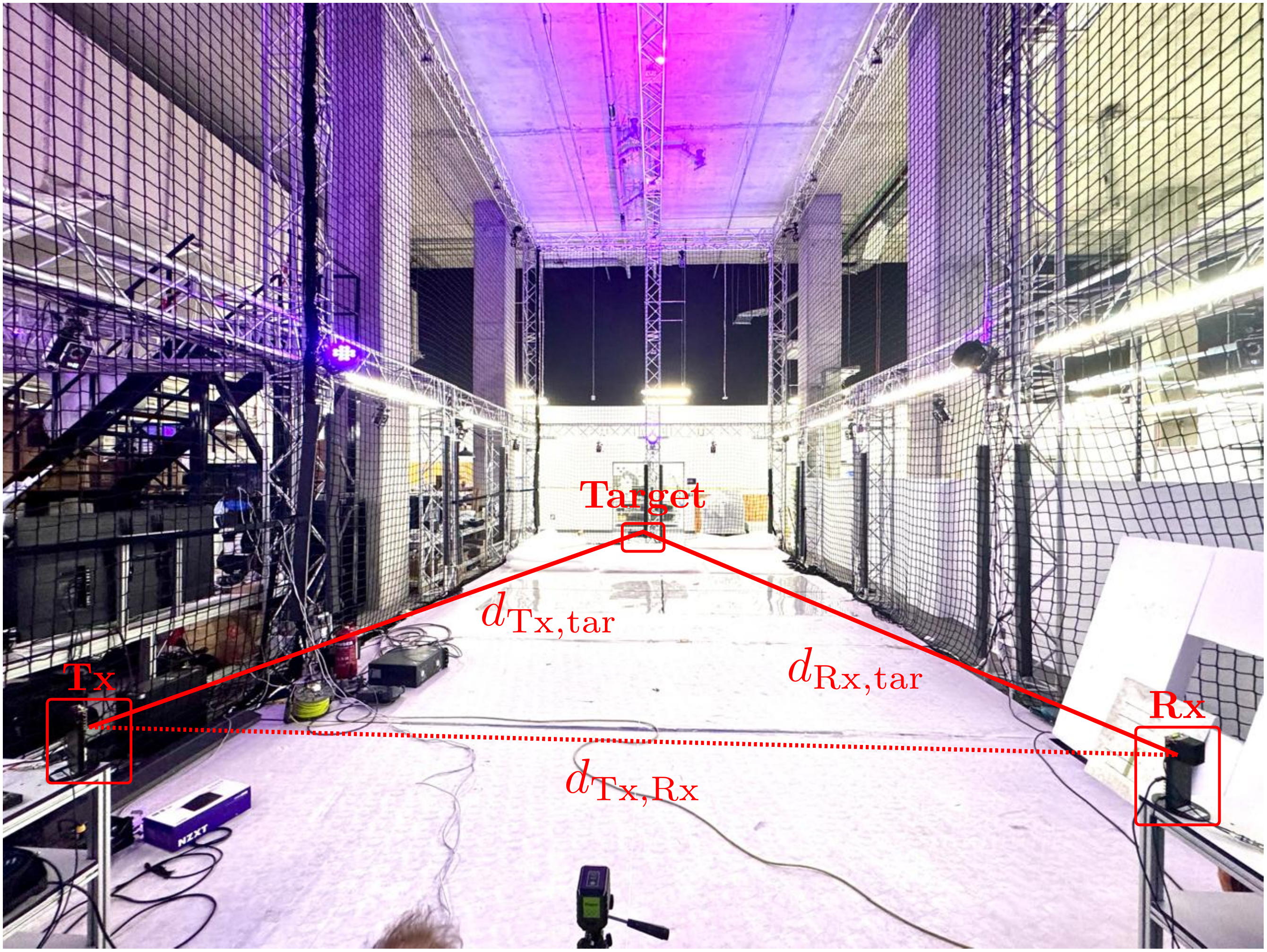} 
    \caption{The {\ac{InF}} environment and the measurement setup.}
    \label{actual_setup} 
\end{figure}

 In this work, we employ monostatic and different bistatic configurations with bistatic angles, \(\theta_b\) of \(20^\circ, 40^\circ\), \(60^\circ\). For the monostatic configuration, \(\theta_b=0\), however, it is impossible to have \(\theta_b=0\) in practice. Therefore, we employ a quasi-monostatic configuration in which the {\ac{Tx}} and {\ac{Rx}} are separated by \(55\) cm, leading to \(\theta_b \approx 10^\circ\). An illustration of the experimental setup is shown in Fig. \ref{measurement_setup}. The distance between the {\ac{Tx}} and the {\ac{Rx}} is given as \(d_\mathrm{Tx, Rx}\), the distance between the {\ac{Tx}} and the target is \(d_\mathrm{Tx,tar}\), and the distance between the Rx and the target is \(d_\mathrm{Rx,tar}\).
 Moreover, the distances, \(d_\mathrm{Tx,tar}\) and \(d_\mathrm{Rx,tar}\) are always the same. The actual measurements are carried out in an {\ac{InF}} environment as shown in Fig. \ref{actual_setup}. The distances \(d_\mathrm{Tx, Rx}\), \(d_\mathrm{Tx,tar}\), and \(d_\mathrm{Rx,tar}\) are also illustrated in Fig. \ref{actual_setup}. The {\ac{Tx}} and the {\ac{Rx}} are placed in the middle of the {\ac{InF}} environment (\(15\times 5\times 8.5\)m). The environment includes the dense clutter characteristic of typical {\ac{InF}} scenarios, ensuring realistic propagation conditions. Moreover, the height of the {\ac{Tx}} and {\ac{Rx}} is \(1\) m from the ground. The operating frequencies that we consider are \(25-28\) GHz. Moreover, {\ac{H}}-polarizations are considered for both {\ac{Tx}} and the {\ac{Rx}}.

The {\ac{Tx}} and {\ac{Rx}} used in these measurements are Sivers \(\mathrm{EVK}02004\) {\acp{RFIC}}, featuring a half-power beamwidth of \(10^\circ\) in both azimuth and elevation planes. Each {\ac{RFIC}} incorporates a \(16\)-element (\(4 \times 4\)) antenna array. The {\ac{Tx}} and {\ac{Rx}} {\acp{RFIC}} are interfaced with \(\mathrm{B205mini}\) {\acp{USRP}}, having a bandwidth of \(20\) MHz. The {\acp{USRP}} operate at an {\ac{IF}} of \(3\) GHz. The {\ac{Tx}} employs a known {\ac{ZC}} sequence of length \(128\) as the transmitted signal, generated using MATLAB. {\ac{ZC}} sequences are specifically chosen due to their constant amplitude and optimal correlation properties. At the {\ac{Rx}}, the signal is processed to extract the {\ac{CIR}}. The {\ac{CIR}} encapsulates information regarding the propagation environment, including the presence or absence of a target, by capturing reflections and scattering characteristics within the channel.

During the preliminary stage of the measurement process, the {\acp{CIR}} are acquired for both monostatic and bistatic configurations for all \(\theta_b\) and operating frequencies in the absence of the target. Reference {\acp{CIR}}, termed as background {\acp{CIR}}, and denoted as \(h_\mathrm{back}(n)\), comprehensively encapsulates the intrinsic contributions of environmental clutter and static scatterers within the measurement domain for different \(\theta_b\) and frequencies. \(h_\mathrm{back}(n)\) serves as a benchmark, facilitating the precise discrimination of target-induced reflections from ambient propagation artifacts in subsequent analysis.

The \acp{CIR}, \(h(n)\) are subsequently estimated in the presence of the target across the configurations above and operating frequencies. \textcolor{black}{The} measurements capture alterations in the propagation environment introduced by the target, \(h(n)\), comprising the contributions by \(h_\mathrm{back}(n)\) and the target-specific components denoted as \(h_\mathrm{tar}(n)\). Firstly, the total power is evaluated as \(P_\mathrm{tot}=\sum\nolimits_n\vert h(n)\vert^2\). \(P_\mathrm{tot}\) contains the power contribution from both the target and the background, i.e., \(P_\mathrm{tot} = P_\mathrm{tar} + P_\mathrm{back} + P_\mathrm{noise}\), where \(P_\mathrm{noise}\) is the noise power. Next, we evaluate the \(P_\mathrm{back}\) as \(P_\mathrm{back} = \sum\nolimits_n \vert h_\mathrm{back}(n)\vert^2\). The target power is then evaluated as \(P_\mathrm{tar} = P_\mathrm{tot} - P_\mathrm{back} - P_\mathrm{noise}\).  \(P_\mathrm{tar}\) represents the power contribution exclusively due to the target-induced scattering and reflections. \(P_\mathrm{tar}\) is obtained for all targets in both monostatic and bistatic configurations for all \(\theta_b\) and operating frequencies, constructing a comprehensive measurement dataset of target-specific power responses across the measurement configuration and frequency space. \(P_\mathrm{tar}\) contains an accurate representation of the target’s {\ac{RCS}}. Considering \(d_\mathrm{Tx,tar} =d_\mathrm{Rx,tar}=d\), \(P_\mathrm{tar}\) can be expressed as:
\begin{equation}\label{radar_equation}
P_\mathrm{tar} = \frac{P_\mathrm{t}G_\mathrm{Tx}G_\mathrm{Rx}\sigma \lambda^2 L}{(4\pi)^3
d^4},
\end{equation}
where \(P_\mathrm{t}\) is the transmit power, \(G_\mathrm{Tx}\) is the {\ac{Tx}} gain, \(G_\mathrm{Rx}\) is the {\ac{Rx}} gain, \(\lambda\) is the wavelength,
\(\sigma\) is the {\ac{RCS}} in m\(^2\), \(L\) are unknown system losses. \textcolor{black}{Also note} that, apart from \(\lambda\), \(d_\mathrm{Tx,tar}\), and \(d_\mathrm{Rx,tar}\), all other parameters in (\ref{radar_equation}) are system-specific. Consequently, a calibration step is crucial to accurately determine these unknown system parameters before reliable {\ac{RCS}} can be obtained. 
\textcolor{black}{The calibration process consists of placing the \ac{Rx} at the target position, where the received power follows the free space \ac{PL} as $P_{\mathrm{Rx}}=\frac{P_{\mathrm{t}} G_{\mathrm{Tx}} G_{\mathrm{Rx}} \lambda^2 L}{(4 \pi)^2 d^2}$, which is computed through extensive measurements, then a system factor quantity is computed as $K(\lambda,d) = \frac{P_{\mathrm{Rx}}}{4\pi d^2}$. The factor $4\pi d^2$ is to match the powers of the single and double \acp{PL}. Finally, during measurement phase, the \ac{RCS} of a test target is obtained as $\sigma = K^{-1}(\lambda,d)P_\mathrm{tar}$.}
%
%
\vspace{-0.25cm}
\subsection{Parametric Distributions and Goodness-of-Fit Tests}\label{sec2c}
To model and analyze the {\ac{RCS}}, various parametric distributions are used to fit the measured data, with each distribution tailored to capture specific statistical characteristics of the measured data. Distributions such as normal, lognormal, Rayleigh, gamma, exponential, and Weibull are systematically considered, using their distinct features to see how these distributions capture the effects of the measured data. The normal distribution is suitable for datasets exhibiting symmetric behavior around the mean. However, for positively skewed data, which is the case with the measured {\ac{RCS}}, the lognormal and Weibull distributions are more effective, as they can capture the asymmetry and heavy-tail behaviors often associated with specular and diffused reflections reflecting from the test targets. The gamma distribution is particularly well suited to modeling a wide range of skewness. \textcolor{black}{Therefore, }it is especially adept at characterizing the measured {\ac{RCS}} data influenced by the mixture of strong and weak reflections. The exponential distribution effectively captures the positively skewed nature of such data, assuming that smaller {\ac{RCS}} values occur more frequently while larger values are exponentially less likely, \textcolor{black}{which can capture scenarios} where the channel is dominated by {\ac{LoS}} reflections. 
\textcolor{black}{Parametric distributions aligned with the measured {\ac{RCS}} data enable the extraction of statistical parameters that not only characterize the data but also reveal nuanced reflective properties of the test targets.} 

The normal distribution is defined by the mean, \(\mu\), and the standard deviation, \(\sigma > 0\). The lognormal distribution uses the mean, \(\mu\) and the standard deviation, \(\sigma \ge 0\) of the natural logarithm of the variable. The gamma distribution has a shape parameter, \(A > 0\), and a scale parameter \(B > 0\), controlling the form and spread. The Weibull distribution is also defined by the shape \(B > 0\) and scale \(A > 0\). The Rayleigh distribution uses a scale parameter \(B > 0\) for spread, and the exponential distribution is based on the rate parameter \(\lambda > 0\), which governs the decay rate.

The {\ac{GoF}} of the parametric distributions is evaluated using the {\ac{KS}} statistic and {\ac{MSE}}, \textcolor{black}{ which offer further insights} into the accuracy of fit between the parametric distribution we consider 
and the measured {\ac{RCS}} data. Moreover, it allows us to determine which distribution can be used to model the {\ac{RCS}} for ISAC channel modeling. 

The {\ac{KS}} statistic measures the maximum absolute deviation between the empirical {\ac{CDF}}, \(F(x)\), and the theoretical {\ac{CDF}} of the fitted model, \(F_{\mathrm{fitted}}(x)\), \textcolor{black}{which is} sensitive to differences in the general shape of the distribution and shows regions where the model cannot mimic important features of the empirical data. A smaller {\ac{KS}} statistic indicates a closer fit between the parametric distribution and the measured data. In contrast, a larger KS statistic value suggests a poor fit. The {\ac{KS}} statistic is expressed as
\begin{equation}\label{ks_stat}
\mathrm{KS~Statistic} = \max \left\vert  F(x) - F_{\mathrm{fitted}}(x) \right\vert.
\end{equation}

The {\ac{MSE}} evaluates the average squared difference between the  empirical {\ac{CDF}} and the fitted {\ac{CDF}} across all data points. Unlike the {\ac{KS}} statistic, which focuses on localized mismatches, {\ac{MSE}} provides a broader {\ac{GoF}} measure by capturing cumulative discrepancies between the theoretical fitted model and the measured data. Smaller {\ac{MSE}} values indicate a closer overall agreement between the empirical and theoretical models, while higher values reflect systematic biases or poor representation of the data’s statistical behavior. {\ac{MSE}} is mathematically defined as:
\begin{equation}\label{mse_stat}
 \mathrm{MSE} = \frac{1}{N} \sum\nolimits_{i=1}^{N} \left( F(x_i) - F_{\mathrm{fitted}}(x_i) \right)^2,
\end{equation}
where \(N\) is the total number of data points.

These metrics help identify the distributions that best capture the statistical characteristics of the measured {\ac{RCS}} data, facilitating precise {\ac{RCS}} modeling of different test targets.
\vspace{-2mm}
\section{RCS Fitting to Parametric Distributions}\label{sec2d}
This subsection presents the fitting of parametric distributions to the measured {\ac{RCS}} data. The analysis involves fitting {\acp{PDF}} and {\acp{CDF}} of the parametric distributions to the measured {\ac{RCS}} data. The theoretical distributions' statistical parameters are optimized to fit the measured {\ac{RCS}} data best. Due to space constraints, we illustrate the {\acp{PDF}} and {\acp{CDF}} of a subset of representative cases, specifically: Mavic \(2\) Pro at \(25\) GHz in a quasi-monostatic configuration,  Matrice \(300\) RTK at \(25\) GHz with  \(\theta_b = 20^\circ\), {\ac{QR}} at \(28\) GHz with \(\theta_b = 40^\circ\), and {\ac{RA}} at \(28\) GHz with \(\theta_b = 60^\circ\). The behavior of the measured {\ac{RCS}} data at other frequencies and \(\theta_b\) demonstrates consistent trends and similar fitting performance to the selected representative cases. To ensure completeness, the fitting parameters for all frequencies and \(\theta_b\) are provided in tabular form for all test targets, facilitating a comprehensive understanding of the statistical characteristics across the different measurement configurations.
\subsubsection{{\ac{RCS}} Distribution Modeling for {\ac{UAV}} Test Cases}
The {\ac{RCS}} distribution for the {\ac{UAV}} test targets comprises two distinct platforms: the Mavic \(2\) Pro and the Matrice \(300\) RTK. \textcolor{black}{Note that} the {\ac{RCS}} measurements were conducted while the {\acp{UAV}} were flying and rotating over the observation point. The rotation of the {\acp{UAV}} facilitated comprehensive {\ac{RCS}} measurements, capturing the scattering profiles from all perspectives, including frontal and lateral views. Consequently, the measured {\ac{RCS}} inherently encapsulates the orientation dependency of the {\ac{UAV}} test targets, ensuring a detailed characterization of the {\ac{RCS}} in the measured data. 
 \begin{figure}[tb]
    \centering
    \includegraphics[width=0.4\textwidth]{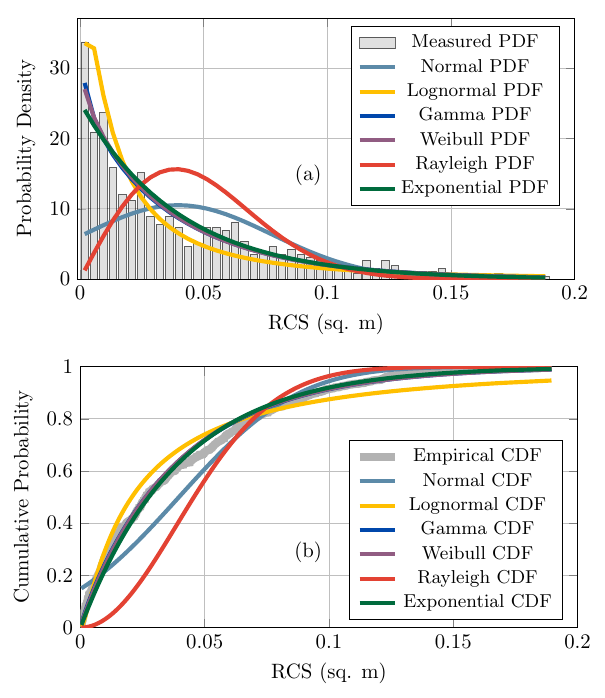} 
    \caption{{\ac{RCS}} Distribution Modeling for Mavic \(2\) Pro at \(25\)GHz for \(\theta_b \approx 10^\circ\): (a) {\ac{PDF}} Fitting, (b) {\ac{CDF}} Fitting. }
    \label{mavic_25_0_results} 
\end{figure}
 \begin{figure}[h]
    \centering
    \includegraphics[width=0.4\textwidth]{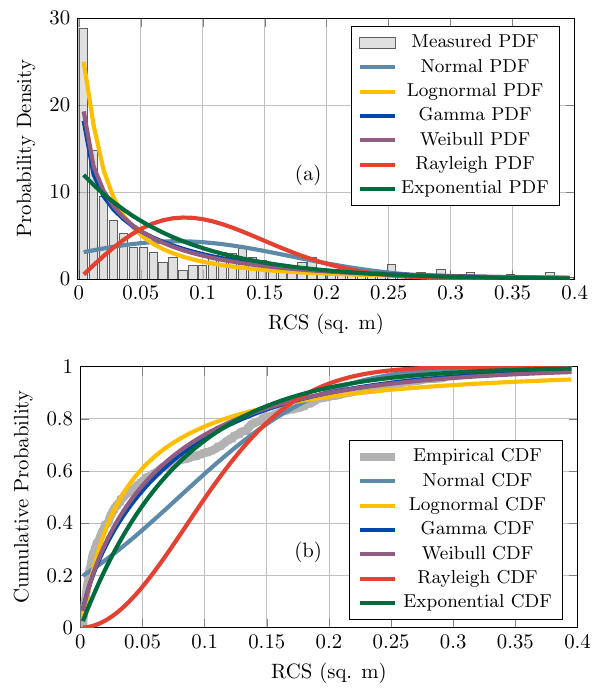} 
    \caption{{\ac{RCS}} Distribution Modeling for Matrice \(300\) RTK at \(25\)GHz for \(\theta_b = 20^\circ\): (a) {\ac{PDF}} Fitting, (b) {\ac{CDF}} Fitting.}
    \label{rtk_25_20_results} 
\end{figure}

Fig. \ref{mavic_25_0_results} illustrates the parametric distribution fitting to the measured {\ac{RCS}} data at \(25\) GHz in a quasi-monostatic configuration for Mavic \(2\) Pro. Fig. \ref{mavic_25_0_results}(a) presents the {\ac{PDF}} fitting, while Fig. \ref{mavic_25_0_results}(b) shows the corresponding {\ac{CDF}} fitting between the measured data and the theoretical distributions. Table \ref{mavic_25} provides a summary of the fitting results. \textcolor{black}{From Fig.} \ref{mavic_25_0_results}(a), \textcolor{black}{we observe that} the measured {\ac{RCS}} data is positively skewed; thus, distributions such as lognormal, Weibull, exponential, and gamma align closely with the measured {\ac{PDF}}, accurately capturing the skewness and tail characteristics. Furthermore, we can also observe that the backscatter reflections from the Mavic \(2\) Pro are generally weak. As detailed in Table \ref{mavic_25}, the gamma distribution provides the best fit based on {\ac{GoF}} metrics. Exponential, Weibull and lognormal distributions also demonstrate good fitting as they capture the skewed nature of the measured {\ac{RCS}} data. However, the normal and Rayleigh distributions show significant deviations, failing to account for the measured {\ac{RCS}} data's asymmetry. Similar conclusions can be drawn from the fitting of the empirical {\ac{CDF}} and the measured {\ac{RCS}} data {\ac{CDF}} as shown in Fig. \ref{mavic_25_0_results}(b), where lognormal, Weibull, exponential, and gamma distributions demonstrate a strong agreement with the empirical {\ac{CDF}}, reinforcing their suitability for modeling the observed {\ac{RCS}} behavior.
\begin{table}[tb]
\centering
\color{black}
\small
\caption{Parameters at \(25\) GHz for varying \(\theta_b\) for Mavic.}
\begin{tabular}{p{1.38cm}|p{1.05cm}|p{1.1cm}|p{3.5cm}}
\hline
\multicolumn{4}{c}{\textbf{25 GHz, \(\theta_b \approx 10^\circ\)}} \\ \hline
\textbf{Dist.} & \textbf{{\ac{KS}} Stat $(\times 10^{-2})$} & \textbf{MSE $(\times 10^{-3})$} & \textbf{Parameters} \\ \hline
Normal & \(15\) & \(6.7\) & \(\mu = 0.04, \sigma = 0.038\) \\
Lognormal & \(8.8\) & \(2.3\) & \(\mu = -3.9, \sigma = 1.4\) \\
Gamma & \(5\) & \(0.43\) & \(A = 0.9, B = 0.044\) \\
Weibull & \(5.2\) & \(0.52\) & \(A = 0.039, B = 0.95\) \\
Rayleigh & \(29\) & \(33\) & \(B = 0.039 \) \\ 
Exponential & \(5.2\) & \(0.87\) & \(\lambda = 0.04\) \\ \hline

\multicolumn{4}{c}{\textbf{25 GHz, \(\theta_b =20^\circ\)}} \\ \hline
Normal & \(16\) & \(7.1\) & \(\mu = 0.05, \sigma = 0.049\) \\
Lognormal & \(13\) & \(7.5\) & \(\mu = -3.90, \sigma = 1.67\) \\
Gamma 	  & \(13\) & \(4.4\) & \(A = 0.66, B = 0.076\) \\
Weibull	 & \(14\) & \(5.1\) & \(A = 0.044, B = 0.77\) \\
Rayleigh & \(33\) & \(35\) & \(B = 0.05\) \\
Exponential & \(20\) & \(9.2\) & \(\lambda = 0.05\) \\ \hline

\multicolumn{4}{c}{\textbf{25 GHz, \(\theta_b =40^\circ\)}} \\ \hline
Normal & \(18\) & \(11\) & \(\mu = 0.046, \sigma = 0.051\) \\
Lognormal & \(10\) & \(2.9\) & \(\mu = -3.92, \sigma = 1.64\) \\
Gamma & \(4.4\) & \(0.22\) & \(A = 0.70, B = 0.066\) \\
Weibull & \(4.2\) & \(0.32\) & \(A = 0.041, B = 0.80\) \\
Rayleigh & \(34\) & \(54\) & \(B = 0.048\) \\
Exponential & \(10\) & \(2.8\) & \(\lambda = 0.046\) \\ \hline

\multicolumn{4}{c}{\textbf{25 GHz, \(\theta_b =60^\circ\)}} \\ \hline
Normal & \(16\) & \(6.3\) & \(\mu = 0.044, \sigma = 0.044\) \\
Lognormal & \(13\) & \(4.4\) & \(\mu = -3.96, \sigma = 1.64\) \\
Gamma & \(8.1\) & \(1.6\) & \(A = 0.71, B = 0.062\) \\
Weibull & \(8.7\) & \(2\) & \(A = 0.04, B = 0.81\) \\
Rayleigh & \(32\) & \(35\) & \(B = 0.044\) \\ 
Exponential & \(13\) & \(4.3\) & \(\lambda = 0.044\) \\ \hline
\end{tabular}
\label{mavic_25}
\end{table}
\begin{table}[h!]
\color{black}
\centering
\small
\caption{Parameters at \(26\) GHz for varying \(\theta_b\) for Mavic.}
\begin{tabular}{p{1.38cm}|p{1.05cm}|p{1.1cm}|p{3.5cm}}
\hline
\multicolumn{4}{c}{\textbf{26 GHz, \(\theta_b \approx 10^\circ\)}} \\ \hline
\textbf{Dist.} & \textbf{{\ac{KS}} Stat $(\times 10^{-2})$} & \textbf{MSE $(\times 10^{-3})$} & \textbf{Parameters} \\ \hline
Normal & \(14\) & \(6.3\) & \(\mu = 0.024, \sigma = 0.014\) \\
Lognormal & \(7.8\) & \(1.4\) & \(\mu = -3.8, \sigma = 0.52\) \\
Gamma & \(9.2\) & \(2.4\) & \(A = 3.66, B = 0.0066\) \\
Weibull & \(12\) & \(3\) & \(A = 0.027, B = 1.86\) \\
Rayleigh & \(11\) & \(4.2\) & \(B = 0.019\) \\ 
Exponential & \(32\) & \(20\) & \(\lambda = 0.024\) \\ \hline

\multicolumn{4}{c}{\textbf{26 GHz, \(\theta_b =20^\circ\)}} \\ \hline
Normal & \(12\) & \(4\) & \(\mu = 0.036, \sigma = 0.031\) \\
Lognormal & \(12\) & \(5.4\) & \(\mu = -3.9, \sigma = 1.48\) \\
Gamma & \(7\) & \(1\) & \(A = 0.98, B = 0.036\) \\
Weibull & \(6\) & \(0.83\) & \(A = 0.036, B = 1.02\) \\
Rayleigh & \(23\) & \(20\) & \(B = 0.034\) \\
Exponential & \(6.8\) & \(1\) & \(\lambda = 0.036\) \\ \hline

\multicolumn{4}{c}{\textbf{26 GHz, \(\theta_b =40^\circ\)}} \\ \hline
Normal & \(34\) & \(42\) & \(\mu = 0.12, \sigma = 0.29\) \\
Lognormal & \(5.4\) & \(1.1\) & \(\mu = -3.9, \sigma = 2.2\) \\
Gamma & \(15\) & \(6.8\) & \(A = 0.35, B = 0.35\) \\
Weibull & \(8.3\) & \(1.7\) & \(A = 0.056, B = 0.4996\) \\
Rayleigh & \(71\) & \(190\) & \(B = 0.22\) \\
Exponential & \(36\) & \(60\) & \(\lambda = 0.12\) \\ \hline

\multicolumn{4}{c}{\textbf{26 GHz, \(\theta_b =60^\circ\)}} \\ \hline
Normal & \(29\) & \(26\) & \(\mu = 0.063, \sigma = 0.11\) \\
Lognormal & \(3.1\) & \(0.2\) & \(\mu = -3.95, \sigma = 1.71\) \\
Gamma & \(9.4\) & \(2.8\) & \(A = 0.52, B = 0.12\) \\
Weibull & \(5.6\) & \(0.86\) & \(A = 0.043, B = 0.64\) \\
Rayleigh & \(57\) & \(130\) & \(B = 0.093\) \\ 
Exponential & \(23\) & \(21\) & \(\lambda = 0.063\) \\ \hline

\end{tabular}
\label{mavic_26}
\end{table}
\begin{table}[h!]
\color{black}
\centering
\small
\caption{Parameters at \(27\) GHz for varying \(\theta_b\) for Mavic.}
\begin{tabular}{p{1.38cm}|p{1.05cm}|p{1.1cm}|p{3.5cm}}
\hline
\multicolumn{4}{c}{\textbf{27 GHz, \(\theta_b \approx 10^\circ\)}} \\ \hline
\textbf{Dist.} & \textbf{{\ac{KS}} Stat}\newline $(\times 10^{-2})$ & \textbf{MSE}\newline$(\times 10^{-3})$ & \textbf{Parameters} \\ \hline
Normal & \(22\) & \(15\) & \(\mu = 0.067, \sigma = 0.088\) \\
Lognormal & \(8.7\) & \(2.7\) & \(\mu = -3.83, \sigma = 1.74\) \\
Gamma & \(9.9\) & \(2.1\) & \(A = 0.55, B = 0.12\) \\
Weibull & \(9.4\) & \(1.8\) & \(A = 0.05, B = 0.66\) \\
Rayleigh & \(44\) & \(78\) & \(B = 0.078\) \\ 
Exponential & \(22\) & \(15\) & \(\lambda = 0.067\) \\ \hline

\multicolumn{4}{c}{\textbf{27 GHz, \(\theta_b =20^\circ\)}} \\ \hline
Normal & \(16\) & \(8.4\) & \(\mu = 0.038, \sigma = 0.039\) \\
Lognormal & \(12\) & \(4.3\) & \(\mu = -3.9, \sigma = 1.46\) \\
Gamma & \(4.5\) & \(4\) & \(A = 0.89, B = 0.043\) \\
Weibull & \(4.1\) & \(3.5\) & \(A = 0.038, B = 0.94\) \\
Rayleigh & \(30\) & \(37\) & \(B = 0.039\) \\
Exponential & \(3\) & \(0.21\) & \(\lambda = 0.038\) \\ \hline

\multicolumn{4}{c}{\textbf{27 GHz, \(\theta_b =40^\circ\)}} \\ \hline
Normal & \(18\) & \(8.3\) & \(\mu = 0.053, \sigma = 0.06\) \\
Lognormal & \(11\) & \(4\) & \(\mu = -3.94, \sigma = 1.80\) \\
Gamma & \(6.2\) & \(1.4\) & \(A = 0.6, B = 0.087\) \\
Weibull & \(7.6\) & \(1.9\) & \(A = 0.044, B = 0.72\) \\
Rayleigh & \(38\) & \(47\) & \(B = 0.056\) \\
Exponential & \(16\) & \(7.5\) & \(\lambda = 0.053\) \\ \hline

\multicolumn{4}{c}{\textbf{27 GHz, \(\theta_b =60^\circ\)}} \\ \hline
Normal & \(25\) & \(21\) & \(\mu = 0.038, \sigma = 0.049\) \\
Lognormal & \(9.3\) & \(3.3\) & \(\mu = -3.97, \sigma = 1.17\) \\
Gamma & \(15\) & \(8.2\) & \(A = 0.82, B = 0.046\) \\
Weibull & \(13\) & \(6\) & \(A = 0.034, B = 0.84\) \\
Rayleigh & \(50\) & \(91\) & \(B = 0.044\) \\ 
Exponential & \(19\) & \(12\) & \(\lambda = 0.038\) \\ \hline

\end{tabular}
\label{mavic_27}
\end{table}
\begin{table}[h!]
\color{black}
\centering
\small
\caption{Parameters at \(28\) GHz for varying \(\theta_b\) for Mavic.}
\begin{tabular}{p{1.38cm}|p{1.05cm}|p{1.1cm}|p{3.5cm}}
\hline
\multicolumn{4}{c}{\textbf{28 GHz, \(\theta_b \approx 10^\circ\)}} \\ \hline
\textbf{Dist.} & \textbf{KS Stat}\newline$(\times 10^{-2})$ & \textbf{MSE}\newline$(\times 10^{-3})$ & \textbf{Parameters} \\ \hline
Normal & \(19\) & \(11\) & \(\mu = 0.027, \sigma = 0.023\) \\
Lognormal & \(6\) & \(0.81\) & \(\mu = -3.79, \sigma = 0.61\) \\
Gamma & \(8.4\) & \(2.4\) & \(A = 2.50, B = 0.011\) \\
Weibull & \(13\) & \(3.8\) & \(A = 0.031, B = 1.42\) \\
Rayleigh & \(21\) & \(20\) & \(B = 0.025\) \\ 
Exponential & \(25\) & \(13\) & \(\lambda = 0.027\) \\ \hline

\multicolumn{4}{c}{\textbf{28 GHz, \(\theta_b =20^\circ\)}} \\ \hline
Normal & \(12\) & \(4.6\) & \(\mu = 0.045, \sigma = 0.04\) \\
Lognormal & \(15\) & \(8.6\) & \(\mu = -3.81, \sigma = 1.77\) \\
Gamma & \(8.6\) & \(1.8\) & \(A = 0.80, B = 0.056\) \\
Weibull & \(7.5\) & \(1.3\) & \(A = 0.044, B = 0.92\) \\
Rayleigh & \(23\) & \(22\) & \(B = 0.043\) \\
Exponential & \(6.2\) & \(8.4\) & \(\lambda = 0.045\) \\ \hline

\multicolumn{4}{c}{\textbf{28 GHz, \(\theta_b =40^\circ\)}} \\ \hline
Normal & \(25\) & \(21\) & \(\mu = 0.027, \sigma = 0.029\) \\
Lognormal & \(9.2\) & \(1.9\) & \(\mu = -3.88, \sigma = 0.69\) \\
Gamma & \(13\) & \(6.1\) & \(A = 1.83, B = 0.015\) \\
Weibull & \(18\) & \(7.3\) & \(A = 0.029, B = 1.21\) \\
Rayleigh & \(34\) & \(54\) & \(B = 0.028\) \\
Exponential & \(25\) & \(10\) & \(\lambda = 0.027\) \\ \hline

\multicolumn{4}{c}{\textbf{28 GHz, \(\theta_b =60^\circ\)}} \\ \hline
Normal & \(20\) & \(15\) & \(\mu = 0.04, \sigma = 0.047\) \\
Lognormal & \(4.5\) & \(0.57\) & \(\mu = -3.91, \sigma = 1.24\) \\
Gamma & \(8.5\) & \(2.4\) & \(A = 0.84, B = 0.047\) \\
Weibull & \(7.5\) & \(1.5\) & \(A = 0.037, B = 0.87\) \\
Rayleigh & \(41\) & \(69\) & \(B = 0.043\) \\ 
Exponential & \(12\) & \(4.9\) & \(\lambda = 0.04\) \\ \hline
\end{tabular}
\label{mavic_28}
\end{table}

Fig. \ref{rtk_25_20_results} illustrates the parametric distribution fitting to the measured {\ac{RCS}} data of the Matrice \(300\) RTK at \(25\)GHz for \(\theta_b = 20^\circ\). \textcolor{black}{Note} that the Matrice \(300\) RTK, with its larger dimensions compared to the Mavic \(2\) Pro, is expected to exhibit a higher {\ac{RCS}}. \textcolor{black}{The increase} in {\ac{RCS}} is attributed to its larger reflective surface area, which contributes to powerful backscatter. Additionally, the lithium-ion battery mounted at the rear of the Matrice \(300\) RTK introduces significant specular reflections, which are foreseen to dominate the scattering profile relative to other structural components of the {\ac{UAV}}, leading to an increase in {\ac{RCS}}. 

The analysis of Fig. \ref{rtk_25_20_results}(a) and (b) reveals an observation that is consistent with that for the Mavic \(2\) Pro, i.e., the {\ac{RCS}} distribution exhibits a positive skewness, characterized by a dominance of low-magnitude reflections and occasional high-magnitude specular reflections that contribute to higher {\ac{RCS}} values. However, the specular reflections are notably more pronounced for the Matrice \(300\) RTK compared to the Mavic \(2\) Pro (cf. Fig. \ref{mavic_25_0_results}(a)). Statistical models, including the lognormal, Weibull, and gamma distributions, effectively capture the underlying characteristics of the {\ac{RCS}}'s {\ac{PDF}}. As detailed in Table \ref{rtk_25}, the Weibull distribution best fits the measured data, followed closely by the gamma and lognormal distributions, which also display good alignment based on {\ac{GoF}} evaluations. The empirical {\ac{CDF}} further supports this observation in Fig. \ref{rtk_25_20_results}(b), where the Weibull, lognormal, and gamma distributions exhibit strong consistency with the empirical {\ac{CDF}}, underscoring their suitability for modeling the observed {\ac{RCS}} characteristics. 

The fitting results for the measured {\ac{RCS}} data of the Mavic \(2\) Pro to various parametric distributions are summarized in Tables \ref{mavic_25}-\ref{mavic_28}, encompassing different \(\theta_b\) and operating frequencies. Similarly, the fitting results for the measured {\ac{RCS}} data of the Matrice \(300\) RTK are detailed in Tables \ref{rtk_25}–\ref{rtk_28} for same \(\theta_b\) and frequencies. For both {\acp{UAV}}, the KS statistics and {\ac{MSE}} were computed for each distribution to evaluate the {\ac{GoF}}. Based on the analysis presented in Tables \ref{mavic_25}-\ref{mavic_28} for the Mavic \(2\) Pro and Tables \ref{rtk_25}-\ref{rtk_28} for the Matrice \(300\) RTK, the following observations can be inferred
\begin{enumerate}
    \item The Weibull, lognormal, and gamma distributions consistently provide the best fit. Specifically, the gamma distribution demonstrates superior performance in \(12\) out of the \(32\) representative cases, encompassing various \(\theta_b\) and operating frequencies. In contrast, the lognormal and Weibull distributions achieve the best fit in \(9\) cases each. Exponential distribution is the best fit in \(2\) cases. However, the differences in the {\ac{GoF}} metrics among these distributions are minimal in most representative cases. Thus, any of these three distributions can effectively model the {\ac{RCS}} of the {\acp{UAV}}. 
    \item Considering the parameters of the best-fit distributions, it is observed that an increase in frequency for a given \(\theta_b\), marginally increases the {\ac{RCS}}.  
    \item For both {\acp{UAV}}, the {\ac{RCS}} decreases with an increase in the \(\theta_b\) for all frequencies. The directional scattering characteristics, i.e., reduced contribution of the specular reflections, the material's and structural properties' angular dependency, etc. may explain the trends observed in the {\ac{RCS}}; however, it is difficult to quantify the relative contribution of each parameter.
    \item Based on the following parameters of the best-fit distributions, it is evident that the {\ac{RCS}} of the Matrice \(300\) RTK is higher compared to that of the Mavic \(2\) Pro. A primary contributing factor is the larger physical dimensions of the Matrice \(300\) RTK. Additionally, the placement of the lithium-ion battery at one end of the Matrice \(300\) RTK results in occasional strong specular reflections, which increases the {\ac{RCS}}. While material properties may also influence the {\ac{RCS}}, quantifying their specific impact remains challenging due to the complexity of their contribution to scattering behavior.
\end{enumerate}
\begin{table}[tb]
\centering
\color{black}
\small
\caption{Parameters at \(25\) GHz for varying \(\theta_b\) for Matrice.}
\begin{tabular}{p{1.38cm}|p{1.05cm}|p{1.1cm}|p{3.5cm}}
\hline
\multicolumn{4}{c}{\textbf{25 GHz, \(\theta_b \approx 10^\circ\)}} \\ \hline
\textbf{Dist.} & \textbf{KS Stat}\newline$(\times 10^{-2})$ & \textbf{MSE}\newline$(\times 10^{-3})$ & \textbf{Parameters} \\ \hline
Normal & \(19\) & \(11\) & \(\mu = 0.063, \sigma = 0.072\) \\
Lognormal & \(10\) & \(2.3\) & \(\mu = -3.5, \sigma = 1.42\) \\
Gamma & \(4.1\) & \(0.4\) & \(A = 0.79, B = 0.079\) \\
Weibull & \(4\) & \(0.4\) & \(A = 0.058, B = 0.86\) \\
Rayleigh & \(33\) & \(54\) & \(B = 0.068\) \\ 
Exponential & \(8.9\) & \(1.8\) & \(\lambda = 0.063\) \\ \hline

\multicolumn{4}{c}{\textbf{25 GHz, \(\theta_b =20^\circ\)}} \\ \hline
Normal & \(20\) & \(13\) & \(\mu = 0.075, \sigma = 0.09\) \\
Lognormal & \(10\) & \(2.2\) & \(\mu = -3.54, \sigma = 1.59\) \\
Gamma & \(8.7\) & \(2.6\) & \(A = 0.63, B = 0.12\) \\
Weibull & \(8.1\) & \(2.2\) & \(A = 0.062, B = 0.73\) \\
Rayleigh & \(44\) & \(64\) & \(B = 0.083\) \\
Exponential & \(19\) & \(12\) & \(\lambda = 0.075\) \\ \hline

\multicolumn{4}{c}{\textbf{25 GHz, \(\theta_b =40^\circ\)}} \\ \hline
Normal & \(19\) & \(12\) & \(\mu = 0.054, \sigma = 0.063\) \\
Lognormal & \(8.6\) & \(2.4\) & \(\mu = -3.59, \sigma = 1.36\) \\
Gamma & \(4.6\) & \(0.32\) & \(A = 0.85, B = 0.063\) \\
Weibull & \(4.2\) & \(0.29\) & \(A = 0.051, B = 0.89\) \\
Rayleigh & \(34\) & \(56\) & \(B = 0.059\) \\
Exponential & \(4.7\) & \(0.79\) & \(\lambda = 0.054\) \\ \hline

\multicolumn{4}{c}{\textbf{25 GHz, \(\theta_b =60^\circ\)}} \\ \hline
Normal & \(20\) & \(13\) & \(\mu = 0.072, \sigma = 0.089\) \\
Lognormal & \(8\) & \(1.7\) & \(\mu = -3.62, \sigma = 1.70\) \\
Gamma & \(4.6\) & \(0.57\) & \(A = 0.61, B = 0.11\) \\
Weibull & \(4.9\) & \(0.53\) & \(A = 0.059, B = 0.72\) \\
Rayleigh & \(41\) & \(67\) & \(B = 0.081\) \\ 
Exponential & \(15\) & \(8.5\) & \(\lambda = 0.072\) \\ \hline

\end{tabular}
\label{rtk_25}
\end{table}
\begin{table}[h!]
\centering
\color{black}
\small
\caption{Parameters at \(26\) GHz for varying \(\theta_b\) for \textcolor{black}{Matrice}.}
\begin{tabular}{p{1.38cm}|p{1.05cm}|p{1.1cm}|p{3.5cm}}
\hline
\multicolumn{4}{c}{\textbf{26 GHz, \(\theta_b \approx 10^\circ\)}} \\ \hline
\textbf{Dist.} & \textbf{{\ac{KS}} Stat}\newline$(\times 10^{-2})$ & \textbf{MSE}\newline$(\times 10^{-3})$ & \textbf{Parameters} \\ \hline
Normal & \(19\) & \(11\) & \(\mu = 0.066, \sigma = 0.074\) \\
Lognormal & \(7.1\) & \(1.3\) & \(\mu = -3.49, \sigma = 1.47\) \\
Gamma & \(3.3\) & \(0.26\) & \(A = 0.77, B = 0.085\) \\
Weibull & \(2.2\) & \(1.3\) & \(A = 0.06, B = 0.84\) \\
Rayleigh & \(38\) & \(57\) & \(B = 0.07\) \\ 
Exponential & \(8.5\) & \(2.6\) & \(\lambda = 0.066\) \\ \hline

\multicolumn{4}{c}{\textbf{26 GHz, \(\theta_b =20^\circ\)}} \\ \hline
Normal & \(17\) & \(8.6\) & \(\mu = 0.097, \sigma = 0.1\) \\
Lognormal & \(15\) & \(7.5\) & \(\mu = -3.53, \sigma = 2.26\) \\
Gamma & \(7.6\) & \(1.6\) & \(A = 0.52, B = 0.18\) \\
Weibull & \(8.4\) & \(2.3\) & \(A = 0.077, B = 0.65\) \\
Rayleigh & \(34\) & \(46\) & \(B = 0.1\) \\
Exponential & \(15\) & \(5.7\) & \(\lambda = 0.097\) \\ \hline

\multicolumn{4}{c}{\textbf{26 GHz, \(\theta_b =40^\circ\)}} \\ \hline
Normal & \(17\) & \(7.2\) & \(\mu = 0.08, \sigma = 0.085\) \\
Lognormal & \(14\) & \(6.5\) & \(\mu = -3.56, \sigma = 2\) \\
Gamma & \(8.1\) & \(1.5\) & \(A = 0.59, B = 0.13\) \\
Weibull & \(8.9\) & \(2\) & \(A = 0.06, B = 0.72\) \\
Rayleigh & \(31\) & \(40\) & \(B = 0.083\) \\
Exponential & \(13\) & \(4.1\) & \(\lambda = 0.08\) \\ \hline

\multicolumn{4}{c}{\textbf{26 GHz, \(\theta_b =60^\circ\)}} \\ \hline
Normal & \(20\) & \(13\) & \(\mu = 0.074, \sigma = 0.089\) \\
Lognormal & \(8.6\) & \(2.3\) & \(\mu = -3.59, \sigma = 1.7462\) \\
Gamma & \(3.4\) & \(0.29\) & \(A = 0.61, B = 0.12\) \\
Weibull & \(4.3\) & \(0.34\) & \(A = 0.061, B = 0.72\) \\
Rayleigh & \(39\) & \(67\) & \(B = 0.082\) \\ 
Exponential & \(14\) & \(6.8\) & \(\lambda = 0.074\) \\ \hline
\end{tabular}
\label{rtk_26}
\end{table}
\begin{table}[h!]
\centering
\color{black}
\small
\caption{Parameters at \(27\) GHz for varying \(\theta_b\) for \textcolor{black}{Matrice}.}
\begin{tabular}{p{1.38cm}|p{1.05cm}|p{1.1cm}|p{3.5cm}}
\hline
\multicolumn{4}{c}{\textbf{27 GHz, \(\theta_b \approx 10^\circ\)}} \\ \hline
\textbf{Dist.} & \textbf{{\ac{KS}} Stat}\newline$(\times 10^{-2})$ & \textbf{MSE}\newline$(\times 10^{-3})$ & \textbf{Parameters} \\ \hline
Normal & \(16\) & \(7.2\) & \(\mu = 0.092, \sigma = 0.093\) \\
Lognormal & \(14\) & \(6\) & \(\mu = -3.47, \sigma = 1.96\) \\
Gamma & \(9.3\) & \(2.9\) & \(A = 0.57, B = 0.16\) \\
Weibull & \(10\) & \(3.7\) & \(A = 0.075, B = 0.69\) \\
Rayleigh & \(35\) & \(38\) & \(B = 0.092\) \\ 
Exponential & \(17\) & \(9.2\) & \(\lambda = 0.092\) \\ \hline

\multicolumn{4}{c}{\textbf{27 GHz, \(\theta_b =20^\circ\)}} \\ \hline
Normal & \(20\) & \(12\) & \(\mu = 0.077, \sigma = 0.092\) \\
Lognormal & \(7.8\) & \(1.4\) & \(\mu = -3.49, \sigma = 1.63\) \\
Gamma & \(5.7\) & \(0.78\) & \(A = 0.64, B = 0.11\) \\
Weibull & \(4.6\) & \(0.67\) & \(A = 0.065, B = 0.74\) \\
Rayleigh & \(41\) & \(63\) & \(B = 0.085\) \\
Exponential & \(15\) & \(7.7\) & \(\lambda = 0.077\) \\ \hline

\multicolumn{4}{c}{\textbf{27 GHz, \(\theta_b =40^\circ\)}} \\ \hline
Normal & \(22\) & \(18\) & \(\mu = 0.057, \sigma = 0.071\) \\
Lognormal & \(3.9\) & \(0.4\) & \(\mu = -3.53, \sigma = 1.19\) \\
Gamma & \(10\) & \(3.5\) & \(A = 0.86, B = 0.065\) \\
Weibull & \(8.6\) & \(2.2\) & \(A = 0.053, B = 0.87\) \\
Rayleigh & \(44\) & \(80\) & \(B = 0.064\) \\
Exponential & \(13\) & \(5.7\) & \(\lambda = 0.057\) \\ \hline

\multicolumn{4}{c}{\textbf{27 GHz, \(\theta_b =60^\circ\)}} \\ \hline
Normal & \(18\) & \(10\) & \(\mu = 0.078, \sigma = 0.089\) \\
Lognormal & \(10\) & \(4\) & \(\mu = -3.58, \sigma = 1.87\) \\
Gamma & \(5.2\) & \(0.5\) & \(A = 0.59, B = 0.13\) \\
Weibull & \(6.3\) & \(0.96\) & \(A = 0.064, B = 0.71\) \\
Rayleigh & \(37\) & \(54\) & \(B = 0.084\) \\ 
Exponential & \(14\) & \(6\) & \(\lambda = 0.078\) \\ \hline
\end{tabular}
\label{rtk_27}
\end{table}
\begin{table}[h!]
\centering
\color{black}
\small
\caption{Parameters at \(28\) GHz for varying \(\theta_b\) for \textcolor{black}{Matrice}.}
\begin{tabular}{p{1.38cm}|p{1.05cm}|p{1.1cm}|p{3.5cm}}
\hline
\multicolumn{4}{c}{\textbf{28 GHz, \(\theta_b \approx 10^\circ\)}} \\ \hline
\textbf{Dist.} & \textbf{KS Stat}\newline$(\times 10^{-2})$ & \textbf{MSE}\newline$(\times 10^{-3})$ & \textbf{Parameters} \\ \hline
Normal & \(18\) & \(10\) & \(\mu = 0.066, \sigma = 0.073\) \\
Lognormal & \(8.2\) & \(1.6\) & \(\mu = -3.49, \sigma = 1.48\) \\
Gamma & \(4.7\) & \(0.41\) & \(A = 0.76, B = 0.087\) \\
Weibull & \(4.3\) & \(0.42\) & \(A = 0.06, B = 0.83\) \\
Rayleigh & \(35\) & \(50\) & \(B = 0.069\) \\ 
Exponential & \(10\) & \(3\) & \(\lambda = 0.066\) \\ \hline

\multicolumn{4}{c}{\textbf{28 GHz, \(\theta_b =20^\circ\)}} \\ \hline
Normal & \(20\) & \(13\) & \(\mu = 0.081, \sigma = 0.097\) \\
Lognormal & \(8.6\) & \(1.8\) & \(\mu = -3.52, \sigma = 1.67\) \\
Gamma & \(7.9\) & \(1.5\) & \(A = 0.607, B = 0.13\) \\
Weibull & \(7\) & \(1.3\) & \(A = 0.065, B = 0.71\) \\
Rayleigh & \(41\) & \(66\) & \(B = 0.089\) \\
Exponential & \(19\) & \(11\) & \(\lambda = 0.081\) \\ \hline

\multicolumn{4}{c}{\textbf{28 GHz, \(\theta_b =40^\circ\)}} \\ \hline
Normal & \(19\) & \(12\) & \(\mu = 0.084, \sigma = 0.099\) \\
Lognormal & \(8.5\) & \(2.3\) & \(\mu = -3.56, \sigma = 1.84\) \\
Gamma & \(5.2\) & \(0.69\) & \(A = 0.57, B = 0.14\) \\
Weibull & \(5.2\) & \(0.87\) & \(A = 0.067, B = 0.69\) \\
Rayleigh & \(41\) & \(63\) & \(B = 0.092\) \\
Exponential & \(17\) & \(10\) & \(\lambda = 0.084\) \\ \hline

\multicolumn{4}{c}{\textbf{28 GHz, \(\theta_b =60^\circ\)}} \\ \hline
Normal & \(19\) & \(11\) & \(\mu = 0.083, \sigma = 0.095\) \\
Lognormal & \(10\) & \(4.9\) & \(\mu = -3.57, \sigma = 2.14\) \\
Gamma & \(3.2\) & \(0.26\) & \(A = 0.57, B = 0.14\) \\
Weibull & \(3.9\) & \(0.43\) & \(A = 0.068, B = 0.7\) \\
Rayleigh & \(37\) & \(58\) & \(B = 0.089\) \\ 
Exponential & \(14\) & \(5.7\) & \(\lambda = 0.083\) \\ \hline
\end{tabular}
\label{rtk_28}
\end{table}
\subsubsection{RCS Distribution Modeling for Robotic Arm}
 \begin{figure}[t]
    \centering
    \includegraphics[width=0.42\textwidth]{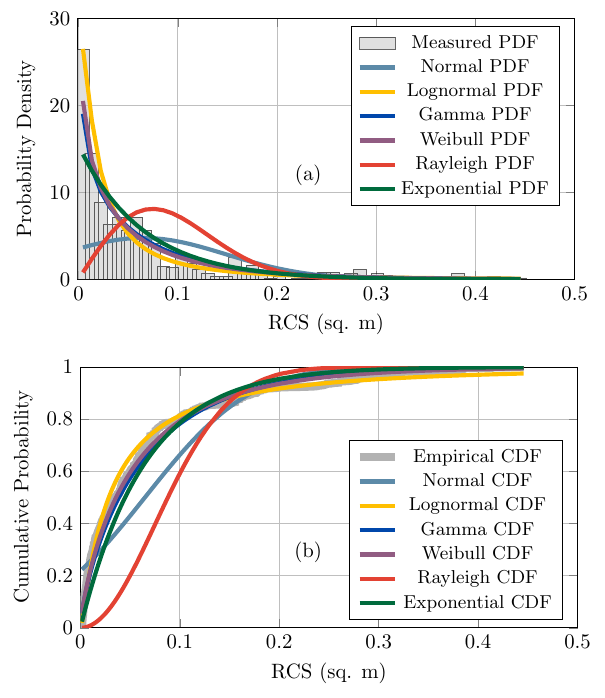} 
    \caption{RCS Distribution Modeling for {\ac{RA}} at \(28\)GHz for \(\theta_b = 60^\circ\): (a) {\ac{PDF}} Fitting, (b) {\ac{CDF}} Fitting.}
    \label{ra_28_60_results} 
\end{figure}
The analysis of the {\ac{RA}}'s RCS data at \(28\) GHz and \(\theta_b = 60^\circ\) is shown in Fig. \ref{ra_28_60_results} which illustrates the fit of different parametric distributions to the measured RCS data. The {\ac{PDF}} in Fig. \ref{ra_28_60_results}(a) reveals that among the fitted statistical models, Weibull, Gamma, and lognormal distributions closely align with the {\ac{PDF}} of the measured RCS data, effectively capturing the nature of backscatter from the {\ac{RA}}. The best fit, however, is provided by Weibull distribution. The fitting results for different distributions are summarized in Table \ref{ra_28}. Models such as the Normal and Rayleigh distributions fail to adequately represent the measured RCS data, as they cannot take into account the impact of a high number of weak reflections and a low number of specular reflections while fitting the measured RCS data highlighting their limitations in modeling asymmetric distributions. The empirical {\ac{CDF}}, as shown in Fig. \ref{ra_28_60_results} (b), further validates these findings. 

The fitting results for the measured RCS data of the {\ac{RA}} to various parametric distributions are presented in Tables \ref{ra_25}-\ref{ra_28}, encompassing all \(\theta_b\) and operating frequencies. The same {\ac{GoF}} metrics employed for the UAVs are utilized to assess the accuracy of each distribution’s fit. From the results summarized in Tables \ref{ra_25}-\ref{ra_28}, the following inferences can be drawn:
\begin{enumerate}
    \item Lognormal, Weibull, and gamma distributions demonstrate the best fit for the measured RCS data of the {\ac{RA}}. Among these, gamma and Weibull distributions each perform optimally in \(7\) of the \(16\) representative cases, spanning various \(\theta_b\) and operating frequencies. The lognormal distribution provides the best fit in \(2\) cases. Similar to the observations for {\acp{UAV}}, the differences in the {\ac{GoF}} metrics among these three distributions are minimal in most cases.
    \item The fitting parameters of the best-fit distributions reveal that an increase in \(\theta_b\) leads to a reduction in the RCS at a fixed operating frequency, \textcolor{black}{which can} be attributed to the same factors previously discussed for {\acp{UAV}}.
    \item It has been observed that the RCS reduces with an increase in frequency for a fixed \(\theta_b\), \textcolor{black}{which is consistent with what has been observed for {\acp{UAV}}.}
    \item The {\ac{RCS}} of {\ac{RA}} is notably higher than that of the {\acp{UAV}}, where \textcolor{black}{the RCS difference is} attributed to its composition, which includes materials that significantly enhance backscatter. In contrast, {\acp{UAV}} predominantly consist of materials such as plastic and carbon fiber, which exhibit lower reflectivity, resulting in reduced RCS values.
\end{enumerate}
\begin{table}[t]
\centering
\color{black}
\small
\caption{Parameters at \(25\) GHz for varying \(\theta_b\) for {\ac{RA}}.}
\begin{tabular}{p{1.38cm}|p{1.05cm}|p{1.1cm}|p{3.5cm}}
\hline
\multicolumn{4}{c}{\textbf{25 GHz, \(\theta_b \approx 10^\circ\)}} \\ \hline
\textbf{Dist.} & \textbf{KS Stat}\newline$(\times 10^{-2})$ & \textbf{MSE}\newline$(\times 10^{-3})$ & \textbf{Parameters} \\ \hline
Normal & \(20\) & \(13\) & \(\mu = 0.085, \sigma = 0.098\) \\
Lognormal & \(11\) & \(3.2\) & \(\mu = -3.43, \sigma = 1.79\) \\
Gamma & \(4.3\) & \(0.34\) & \(A = 0.63, B = 0.13\) \\
Weibull & \(4.5\) & \(0.45\) & \(A = 0.072, B = 0.74\) \\
Rayleigh & \(39\) & \(63\) & \(B = 0.092\) \\ 
Exponential & \(14\) & \(5\) & \(\lambda = 0.085\) \\ \hline

\multicolumn{4}{c}{\textbf{25 GHz, \(\theta_b =20^\circ\)}} \\ \hline
Normal & \(18\) & \(9.4\) & \(\mu = 0.10, \sigma = 0.11\) \\
Lognormal & \(11\) & \(4.7\) & \(\mu = -3.48, \sigma = 2.23\) \\
Gamma & \(8.4\) & \(1.1\) & \(A = 0.52, B = 0.19\) \\
Weibull & \(9.7\) & \(1.6\) & \(A = 0.08, B = 0.65\) \\
Rayleigh & \(39\) & \(50\) & \(B = 0.1\) \\
Exponential & \(18\) & \(9.5\) & \(\lambda = 0.1\) \\ \hline

\multicolumn{4}{c}{\textbf{25 GHz, \(\theta_b =40^\circ\)}} \\ \hline
Normal & \(11\) & \(3.3\) & \(\mu = 0.04, \sigma = 0.03\) \\
Lognormal & \(10\) & \(3.1\) & \(\mu = -3.5, \sigma = 1.08\) \\
Gamma & \(7.1\) & \(1.1\) & \(A = 1.29, B = 0.03\) \\
Weibull & \(6.3\) & \(1.1\) & \(A = 0.049, B = 1.21\) \\
Rayleigh & \(22\) & \(14\) & \(B = 0.041\) \\
Exponential & \(8.2\) & \(1.9\) & \(\lambda = 0.046\) \\ \hline

\multicolumn{4}{c}{\textbf{25 GHz, \(\theta_b =60^\circ\)}} \\ \hline
Normal & \(16\) & \(7.5\) & \(\mu = 0.09, \sigma = 0.09\) \\
Lognormal & \(15\) & \(9.5\) & \(\mu = -3.52, \sigma = 2.35\) \\
Gamma & \(8\) & \(1.9\) & \(A = 0.55, B = 0.16\) \\
Weibull & \(8.4\) & \(2\) & \(A = 0.076, B = 0.69\) \\
Rayleigh & \(32\) & \(40\) & \(B = 0.093\) \\ 
Exponential & \(11\) & \(3.5\) & \(\lambda = 0.091\) \\ \hline

\end{tabular}
\label{ra_25}
\end{table}
\begin{table}[h!]
\centering
\small
\color{black}
\caption{Parameters at \(26\) GHz for varying \(\theta_b\) for {\ac{RA}}.}
\begin{tabular}{p{1.38cm}|p{1.05cm}|p{1.1cm}|p{3.5cm}}
\hline
\multicolumn{4}{c}{\textbf{26 GHz, \(\theta_b \approx 10^\circ\)}} \\ \hline
\textbf{Dist.} & \textbf{KS Stat}\newline$(\times 10^{-2})$ & \textbf{MSE}\newline$(\times 10^{-3})$ & \textbf{Parameters} \\ \hline
Normal & \(20\) & \(13\) & \(\mu = 0.08, \sigma = 0.09\) \\
Lognormal & \(11\) & \(3.4\) & \(\mu = -3.45, \sigma = 1.83\) \\
Gamma & \(4.3\) & \(0.34\) & \(A = 0.62, B = 0.13\) \\
Weibull & \(4.4\) & \(0.46\) & \(A = 0.07, B = 0.73\) \\
Rayleigh & \(39\) & \(63\) & \(B = 0.091\) \\ 
Exponential & \(14\) & \(5.1\) & \(\lambda = 0.084\) \\ \hline

\multicolumn{4}{c}{\textbf{26 GHz, \(\theta_b =20^\circ\)}} \\ \hline
Normal & \(15\) & \(6.2\) & \(\mu = 0.058, \sigma = 0.058\) \\
Lognormal & \(10\) & \(3.1\) & \(\mu = -3.49, \sigma = 1.41\) \\
Gamma & \(5.3\) & \(0.51\) & \(A = 0.88, B = 0.066\) \\
Weibull & \(5.4\) & \(0.57\) & \(A = 0.057, B = 0.93\) \\
Rayleigh & \(30\) & \(33\) & \(B = 0.058\) \\
Exponential & \(5.6\) & \(0.76\) & \(\lambda = 0.058\) \\ \hline

\multicolumn{4}{c}{\textbf{26 GHz, \(\theta_b =40^\circ\)}} \\ \hline
Normal & \(21\) & \(15\) & \(\mu = 0.08, \sigma = 0.1\) \\
Lognormal & \(10\) & \(1.8\) & \(\mu = -3.54, \sigma = 1.64\) \\
Gamma & \(8.9\) & \(2.3\) & \(A = 0.59, B = 0.13\) \\
Weibull & \(7.5\) & \(1.8\) & \(A = 0.064, B = 0.7\) \\
Rayleigh & \(47\) & \(73\) & \(B = 0.092\) \\
Exponential & \(20\) & \(14\) & \(\lambda = 0.08\) \\ \hline

\multicolumn{4}{c}{\textbf{26 GHz, \(\theta_b =60^\circ\)}} \\ \hline
Normal & \(25\) & \(20\) & \(\mu = 0.069, \sigma = 0.089\) \\
Lognormal & \(5.1\) & \(0.62\) & \(\mu = -3.55, \sigma = 1.47\) \\
Gamma & \(11\) & \(3.8\) & \(A = 0.68, B = 0.1\) \\
Weibull & \(9.7\) & \(2.4\) & \(A = 0.058, B = 0.76\) \\
Rayleigh & \(48\) & \(91\) & \(B = 0.08\) \\ 
Exponential & \(20\) & \(12\) & \(\lambda = 0.069\) \\ \hline
\end{tabular}
\label{ra_26}
\end{table}
\begin{table}[h!]
\centering
\small
\color{black}
\caption{Parameters at \(27\) GHz for varying \(\theta_b\) for {\ac{RA}}.}
\begin{tabular}{p{1.38cm}|p{1.05cm}|p{1.1cm}|p{3.5cm}}
\hline
\multicolumn{4}{c}{\textbf{27 GHz, \(\theta_b \approx 10^\circ\)}} \\ \hline
\textbf{Dist.} & \textbf{KS Stat}\newline$(\times 10^{-2})$ & \textbf{MSE}\newline$(\times 10^{-3})$ & \textbf{Parameters} \\ \hline
Normal & \(20\) & \(12\) & \(\mu = 0.089, \sigma = 0.1\) \\
Lognormal & \(8.5\) & \(2.6\) & \(\mu = -3.48, \sigma = 1.82\) \\
Gamma & \(6.9\) & \(0.96\) & \(A = 0.5820, B = 0.15\) \\
Weibull & \(7.1\) & \(1.1\) & \(A = 0.072, B = 0.69\) \\
Rayleigh & \(42\) & \(62\) & \(B = 0.095\) \\ 
Exponential & \(16\) & \(10\) & \(\lambda = 0.089\) \\ \hline

\multicolumn{4}{c}{\textbf{27 GHz, \(\theta_b =20^\circ\)}} \\ \hline
Normal & \(21\) & \(13\) & \(\mu = 0.083, \sigma = 0.10\) \\
Lognormal & \(9.2\) & \(2\) & \(\mu = -3.48, \sigma = 1.75\) \\
Gamma & \(4.3\) & \(0.32\) & \(A = 0.61, B = 0.13\) \\
Weibull & \(3.4\) & \(0.3\) & \(A = 0.068, B = 0.72\) \\
Rayleigh & \(40\) & \(69\) & \(B = 0.093\) \\
Exponential & \(14\) & \(6.8\) & \(\lambda = 0.083\) \\ \hline

\multicolumn{4}{c}{\textbf{27 GHz, \(\theta_b =40^\circ\)}} \\ \hline
Normal & \(13\) & \(7.4\) & \(\mu = 0.032, \sigma = 0.015\) \\
Lognormal & \(6.1\) & \(0.6\) & \(\mu = -3.52, \sigma = 0.38\) \\
Gamma & \(7.5\) & \(1.8\) & \(A = 6.18, B = 0.0052\) \\
Weibull & \(14\) & \(6\) & \(A = 0.036, B = 2.17\) \\
Rayleigh & \(16\) & \(7.1\) & \(B = 0.025\) \\
Exponential & \(37\) & \(37\) & \(\lambda = 0.032\) \\ \hline

\multicolumn{4}{c}{\textbf{27 GHz, \(\theta_b =60^\circ\)}} \\ \hline
Normal & \(21\) & \(14\) & \(\mu = 0.074, \sigma = 0.089\) \\
Lognormal & \(6.1\) & \(0.94\) & \(\mu = -3.54, \sigma = 1.59\) \\
Gamma & \(6.9\) & \(1.3\) & \(A = 0.64, B = 0.11\) \\
Weibull & \(6.1\) & \(0.89\) & \(A = 0.062, B = 0.74\) \\
Rayleigh & \(45\) & \(71\) & \(B = 0.082\) \\ 
Exponential & \(16\) & \(9.5\) & \(\lambda = 0.074\) \\ \hline
\end{tabular}
\label{ra_27}
\end{table}
\begin{table}[h!]
\centering
\small
\color{black}
\caption{Parameters at \(28\) GHz for varying \(\theta_b\) for {\ac{RA}}.}
\begin{tabular}{p{1.38cm}|p{1.05cm}|p{1.1cm}|p{3.5cm}}
\hline
\multicolumn{4}{c}{\textbf{28 GHz, \(\theta_b \approx 10^\circ\)}} \\ \hline
\textbf{Dist.} & \textbf{KS Stat}\newline$(\times 10^{-2})$ & \textbf{MSE}\newline$(\times 10^{-3})$ & \textbf{Parameters} \\ \hline
Normal & \(15\) & \(7\) & \(\mu = 0.069, \sigma = 0.069\) \\
Lognormal & \(13\) & \(5\) & \(\mu = -3.49, \sigma = 1.67\) \\
Gamma & \(5.8\) & \(0.78\) & \(A = 0.73, B = 0.094\) \\
Weibull & \(5.8\) & \(0.99\) & \(A = 0.063, B = 0.83\) \\
Rayleigh & \(30\) & \(36\) & \(B = 0.069\) \\ 
Exponential & \(8.7\) & \(1.7\) & \(\lambda = 0.069\) \\ \hline

\multicolumn{4}{c}{\textbf{28 GHz, \(\theta_b =20^\circ\)}} \\ \hline
Normal & \(19\) & \(11\) & \(\mu = 0.062, \sigma = 0.072\) \\
Lognormal & \(8.5\) & \(1.7\) & \(\mu = -3.52, \sigma = 1.39\) \\
Gamma & \(6.3\) & \(0.68\) & \(A = 0.78, B = 0.079\) \\
Weibull & \(5.7\) & \(0.56\) & \(A = 0.057, B = 0.84\) \\
Rayleigh & \(35\) & \(55\) & \(B = 0.067\) \\
Exponential & \(11\) & \(3\) & \(\lambda = 0.062\) \\ \hline

\multicolumn{4}{c}{\textbf{28 GHz, \(\theta_b =40^\circ\)}} \\ \hline
Normal & \(17\) & \(8.8\) & \(\mu = 0.09, \sigma = 0.095\) \\
Lognormal & \(13\) & \(6\) & \(\mu = -3.54, \sigma = 2.24\) \\
Gamma & \(7.7\) & \(1.2\) & \(A = 0.54, B = 0.16\) \\
Weibull & \(8.7\) & \(1.7\) & \(A = 0.07, B = 0.68\) \\
Rayleigh & \(37\) & \(45\) & \(B = 0.09\) \\
Exponential & \(16\) & \(6.8\) & \(\lambda = 0.09\) \\ \hline

\multicolumn{4}{c}{\textbf{28 GHz, \(\theta_b =60^\circ\)}} \\ \hline
Normal & \(22\) & \(16\) & \(\mu = 0.064, \sigma = 0.083\) \\
Lognormal & \(7\) & \(1\) & \(\mu = -3.55, \sigma = 1.40\) \\
Gamma & \(6.5\) & \(1.3\) & \(A = 0.73, B = 0.08\) \\
Weibull & \(5.4\) & \(0.67\) & \(A = 0.056, B = 0.79\) \\
Rayleigh & \(41\) & \(78\) & \(B = 0.074\) \\ 
Exponential & \(13\) & \(5.9\) & \(\lambda = 0.064\) \\ \hline
\end{tabular}
\label{ra_28}
\end{table}
\subsubsection{RCS Distribution Modeling for Quadruped Robot}
 \begin{figure}[th]
    \centering
    \includegraphics[width=0.42\textwidth]{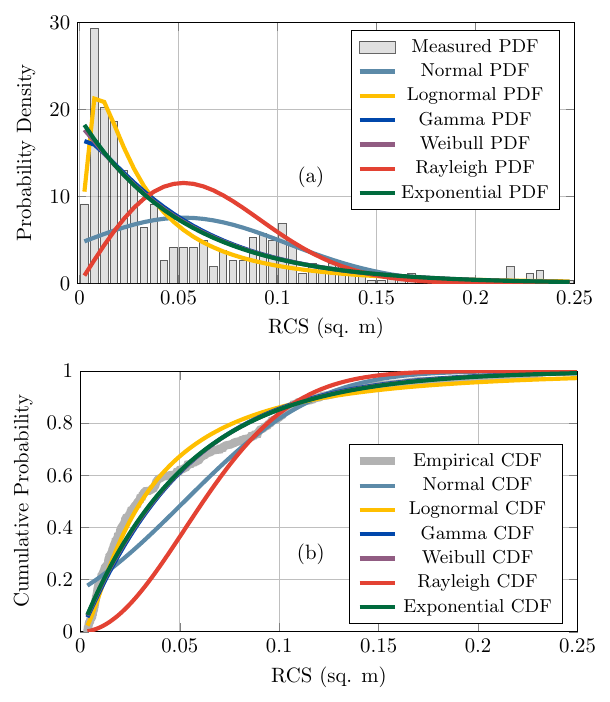} 
    \caption{RCS Distribution Modeling for {\ac{QR}} at \(28\)GHz for \(\theta_b = 40^\circ\): (a) {\ac{PDF}} Fitting, (b) {\ac{CDF}} Fitting.}
    \label{robot_28_40_results} 
\end{figure}
 \begin{figure}[th]
    \centering
    \includegraphics[width=0.42\textwidth]{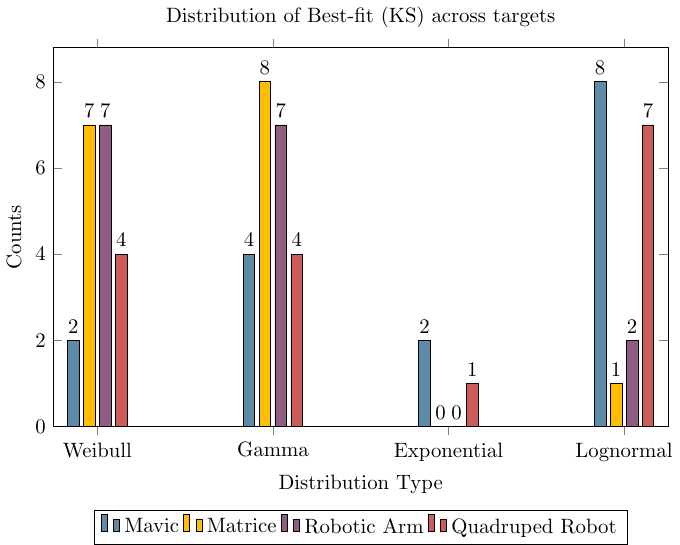} 
    \caption{\textcolor{black}{Unified selected distributions over $25$-$28$ GHz and all bistatic angles.}}
    \label{summary_statistical} 
\end{figure}
The fitting of the measured {\ac{RCS}} data for a {\ac{QR}} at \(28\) GHz and \(\theta_b = 40^\circ\) is depicted in Fig. \ref{robot_28_40_results}. The measured {\ac{RCS}} data exhibits a positively skewed {\ac{PDF}}, characterized by a predominance of weak reflections and a limited number of strong backscatter reflections from the target, \textcolor{black}{which is anticipated because} the continuous movement of the {\ac{QR}} introduces orientation-dependent variations in the {\ac{RCS}}. Consequently, strong reflections are observed when the robot’s orientation favors specular backscatter, while weak reflections occur when the orientation reduces effective backscatter.

Fig. \ref{robot_28_40_results}(a) indicates that the lognormal distribution provides the closest fit to the measured {\ac{RCS}} data. \textcolor{black}{Our} findings are further corroborated by the {\ac{GoF}} metrics summarized in Table \ref{robot_28}. The Weibull and Exponential distributions show moderate alignment. The empirical {\ac{CDF}} in Fig. \ref{robot_28_40_results}(b) also validates these results, demonstrating strong agreement between the empirical data and the fitted lognormal and Weibull distributions. The measured {\ac{RCS}} data of the {\ac{QR}}, fitted to various parametric distributions, is summarized in Tables \ref{robot_25}-\ref{robot_28}, covering all considered \(\theta_b\) and operating frequencies. The {\ac{GoF}} metrics evaluate each distribution’s accuracy. The following conclusions can be drawn from the results:
\begin{enumerate}
    \item Lognormal distribution provides the best fit with the measured {\ac{RCS}} data in most of the representative cases. Among these, the lognormal distribution performs best in \(7\) out of \(16\) representative cases, transiting various \(\theta_b\) and operating frequencies. The Weibull, gamma, and exponential distributions best fit in \(4\), \(4\), and \(1\) cases, respectively. \textcolor{black}{The differences in terms of} {\ac{GoF}} metrics between lognormal, Weibull and gamma distributions are minimal in most cases.
    \item It is observed that the {\ac{RCS}} decreases with an increase in the bistatic angle across all considered frequencies. Furthermore, the {\ac{RCS}} slightly increases with an increase in the operating frequency for a given \(\theta_b\).
\end{enumerate}

\begin{table}[h!]
\centering
\small
\color{black}
\caption{Parameters at \(25\) GHz for varying \(\theta_b\) for {\ac{QR}}.}
\begin{tabular}{p{1.38cm}|p{1.05cm}|p{1.1cm}|p{3.5cm}}
\hline
\multicolumn{4}{c}{\textbf{25 GHz, \(\theta_b \approx 10^\circ\)}} \\ \hline
\textbf{Dist.} & \textbf{KS Stat}\newline$(\times 10^{-2})$ & \textbf{MSE}\newline$(\times 10^{-3})$ & \textbf{Parameters} \\ \hline
Normal & \(18\) & \(10\) & \(\mu = 0.072, \sigma = 0.077\) \\
Lognormal & \(9.9\) & \(2.1\) & \(\mu = -3.44, \sigma = 1.52\) \\
Gamma & \(5.1\) & \(0.8\) & \(A = 0.73, B = 0.099\) \\
Weibull & \(5.7\) & \(0.87\) & \(A = 0.065, B = 0.81\) \\
Rayleigh & \(38\) & \(49\) & \(B = 0.074\) \\ 
Exponential & \(10\) & \(4.3\) & \(\lambda = 0.07\) \\ \hline

\multicolumn{4}{c}{\textbf{25 GHz, \(\theta_b =20^\circ\)}} \\ \hline
Normal & \(17\) & \(8.6\) & \(\mu = 0.055, \sigma = 0.054\) \\
Lognormal & \(11\) & \(2.6\) & \(\mu = -3.48, \sigma = 1.17\) \\
Gamma & \(9.3\) & \(2.7\) & \(A = 0.98, B = 0.056\) \\
Weibull & \(8.9\) & \(2.5\) & \(A = 0.054, B = 0.98\) \\
Rayleigh & \(35\) & \(39\) & \(B = 0.054\) \\
Exponential & \(9.6\) & \(2.8\) & \(\lambda = 0.055\) \\ \hline

\multicolumn{4}{c}{\textbf{25 GHz, \(\theta_b =40^\circ\)}} \\ \hline
Normal & \(17\) & \(11\) & \(\mu = 0.053, \sigma = 0.054\) \\
Lognormal & \(5.2\) & \(0.86\) & \(\mu = -3.50, \sigma = 1.18\) \\
Gamma & \(5.9\) & \(0.86\) & \(A = 0.99, B = 0.054\) \\
Weibull & \(5.8\) & \(0.69\) & \(A = 0.053, B = 0.98\) \\
Rayleigh & \(32\) & \(47\) & \(B = 0.054\) \\
Exponential & \(6\) & \(0.88\) & \(\lambda = 0.053\) \\ \hline

\multicolumn{4}{c}{\textbf{25 GHz, \(\theta_b =60^\circ\)}} \\ \hline
Normal & \(19\) & \(13\) & \(\mu = 0.046, \sigma = 0.048\) \\
Lognormal & \(4.4\) & \(0.5\) & \(\mu = -3.52, \sigma = 0.98\) \\
Gamma & \(7.1\) & \(1.4\) & \(A = 1.22, B = 0.038\) \\
Weibull & \(7.3\) & \(1.3\) & \(A = 0.048, B = 1.07\) \\
Rayleigh & \(30\) & \(46\) & \(B = 0.047\) \\ 
Exponential & \(8.3\) & \(1.1\) & \(\lambda = 0.046\) \\ \hline
\end{tabular}
\label{robot_25}
\end{table}
\begin{table}[h!]
\centering
\small
\color{black}
\caption{Parameters at \(26\) GHz for varying \(\theta_b\) for {\ac{QR}}.}
\begin{tabular}{p{1.38cm}|p{1.05cm}|p{1.1cm}|p{3.5cm}}
\hline
\multicolumn{4}{c}{\textbf{26 GHz, \(\theta_b \approx 10^\circ\)}} \\ \hline
\textbf{Dist.} & \textbf{KS Stat}\newline$(\times 10^{-2})$ & \textbf{MSE}\newline$(\times 10^{-3})$ & \textbf{Parameters} \\ \hline
Normal & \(18\) & \(10\) & \(\mu = 0.055, \sigma = 0.056\) \\
Lognormal & \(9.9\) & \(2\) & \(\mu = -3.44, \sigma = 1.10\) \\
Gamma & \(11\) & \(2.1\) & \(A = 1.04, B = 0.053\) \\
Weibull & \(10\) & \(1.7\) & \(A = 0.055, B = 0.99\) \\
Rayleigh & \(33\) & \(44\) & \(B = 0.056\) \\ 
Exponential & \(10\) & \(1.7\) & \(\lambda = 0.055\) \\ \hline

\multicolumn{4}{c}{\textbf{26 GHz, \(\theta_b =20^\circ\)}} \\ \hline
Normal & \(15\) & \(6.2\) & \(\mu = 0.058, \sigma = 0.055\) \\
Lognormal & \(10\) & \(3\) & \(\mu = -3.46, \sigma = 1.3\) \\
Gamma & \(7.9\) & \(1.3\) & \(A = 0.92, B = 0.063\) \\
Weibull & \(8.2\) & \(1.5\) & \(A = 0.057, B = 0.95\) \\
Rayleigh & \(32\) & \(32\) & \(B = 0.057\) \\
Exponential & \(9.7\) & \(1.9\) & \(\lambda = 0.058\) \\ \hline

\multicolumn{4}{c}{\textbf{26 GHz, \(\theta_b =40^\circ\)}} \\ \hline
Normal & \(14\) & \(5.3\) & \(\mu = 0.058, \sigma = 0.055\) \\
Lognormal & \(11\) & \(3.8\) & \(\mu = -3.5, \sigma = 1.4\) \\
Gamma & \(6.6\) & \(0.97\) & \(A = 0.87, B = 0.066\) \\
Weibull & \(6.7\) & \(1.1\) & \(A = 0.057, B = 0.94\) \\
Rayleigh & \(28\) & \(28\) & \(B = 0.057\) \\
Exponential & \(6.4\) & \(1.3\) & \(\lambda = 0.058\) \\ \hline

\multicolumn{4}{c}{\textbf{26 GHz, \(\theta_b =60^\circ\)}} \\ \hline
Normal & \(20\) & \(14\) & \(\mu = 0.052, \sigma = 0.056\) \\
Lognormal & \(5.7\) & \(1.1\) & \(\mu = -3.55, \sigma = 1.14\) \\
Gamma & \(10\) & \(3.2\) & \(A = 0.96, B = 0.054\) \\
Weibull & \(8.6\) & \(2.4\) & \(A = 0.05, B = 0.94\) \\
Rayleigh & \(39\) & \(60\) & \(B = 0.054\) \\ 
Exponential & \(10\) & \(3.7\) & \(\lambda = 0.052\) \\ \hline
\end{tabular}
\label{robot_26}
\end{table}
\begin{table}[h!]
\centering
\color{black}
\small
\caption{Parameters at \(27\) GHz for varying \(\theta_b\) for {\ac{QR}}.}
\begin{tabular}{p{1.38cm}|p{1.05cm}|p{1.1cm}|p{3.5cm}}
\hline
\multicolumn{4}{c}{\textbf{27 GHz, \(\theta_b \approx 10^\circ\)}} \\ \hline
\textbf{Dist.} & \textbf{KS Stat}\newline$(\times 10^{-2})$ & \textbf{MSE}\newline$(\times 10^{-3})$ & \textbf{Parameters} \\ \hline
Normal & \(16\) & \(7.1\) & \(\mu = 0.062, \sigma = 0.062\) \\
Lognormal & \(14\) & \(4.2\) & \(\mu = -3.42, \sigma = 1.3\) \\
Gamma & \(9.2\) & \(2.2\) & \(A = 0.89, B = 0.069\) \\
Weibull & \(9.1\) & \(2.3\) & \(A = 0.06, B = 0.93\) \\
Rayleigh & \(32\) & \(33\) & \(B = 0.062\) \\ 
Exponential & \(11\) & \(3\) & \(\lambda = 0.062\) \\ \hline

\multicolumn{4}{c}{\textbf{27 GHz, \(\theta_b =20^\circ\)}} \\ \hline
Normal & \(15\) & \(8.5\) & \(\mu = 0.069, \sigma = 0.068\) \\
Lognormal & \(11\) & \(2.9\) & \(\mu = -3.45, \sigma = 1.53\) \\
Gamma & \(7.9\) & \(1.3\) & \(A = 0.75, B = 0.092\) \\
Weibull & \(8.4\) & \(1.6\) & \(A = 0.064, B = 0.84\) \\
Rayleigh & \(33\) & \(40\) & \(B = 0.069\) \\
Exponential & \(14\) & \(4.1\) & \(\lambda = 0.069\) \\ \hline

\multicolumn{4}{c}{\textbf{27 GHz, \(\theta_b =40^\circ\)}} \\ \hline
Normal & \(14\) & \(5\) & \(\mu = 0.056, \sigma = 0.052\) \\
Lognormal & \(10\) & \(3.6\) & \(\mu = -3.47, \sigma = 1.29\) \\
Gamma & \(6.9\) & \(1.6\) & \(A = 0.95, B = 0.059\) \\
Weibull & \(7.1\) & \(1.7\) & \(A = 0.056, B = 0.98\) \\
Rayleigh & \(29\) & \(26\) & \(B = 0.054\) \\
Exponential & \(7.6\) & \(1.8\) & \(\lambda = 0.056\) \\ \hline

\multicolumn{4}{c}{\textbf{27 GHz, \(\theta_b =60^\circ\)}} \\ \hline
Normal & \(22\) & \(17\) & \(\mu = 0.059, \sigma = 0.067\) \\
Lognormal & \(10\) & \(3.8\) & \(\mu = -3.49, \sigma = 1.16\) \\
Gamma & \(15\) & \(7\) & \(A = 0.88, B = 0.067\) \\
Weibull & \(13\) & \(5.5\) & \(A = 0.055, B = 0.88\) \\
Rayleigh & \(45\) & \(72\) & \(B = 0.063\) \\ 
Exponential & \(18\) & \(9.6\) & \(\lambda = 0.059\) \\ \hline
\end{tabular}
\label{robot_27}
\end{table}
\begin{table}[h!]
\centering
\color{black}
\small
\caption{Parameters at \(28\) GHz for varying \(\theta_b\) for {\ac{QR}}.}
\begin{tabular}{p{1.38cm}|p{1.05cm}|p{1.1cm}|p{3.5cm}}
\hline
\multicolumn{4}{c}{\textbf{28 GHz, \(\theta_b \approx 10^\circ\)}} \\ \hline
\textbf{Dist.} & \textbf{KS Stat}\newline$(\times 10^{-2})$ & \textbf{MSE}\newline$(\times 10^{-3})$ & \textbf{Parameters} \\ \hline
Normal & \(15\) & \(7\) & \(\mu = 0.06, \sigma = 0.057\) \\
Lognormal & \(8.8\) & \(2\) & \(\mu = -3.4, \sigma = 1.22\) \\
Gamma & \(6.7\) & \(1.2\) & \(A = 0.97, B = 0.061\) \\
Weibull & \(6.6\) & \(1.2\) & \(A = 0.059, B = 0.98\) \\
Rayleigh & \(32\) & \(34\) & \(B = 0.058\) \\ 
Exponential & \(7.2\) & \(1.4\) & \(\lambda = 0.06\) \\ \hline

\multicolumn{4}{c}{\textbf{28 GHz, \(\theta_b =20^\circ\)}} \\ \hline
Normal & \(18\) & \(9.9\) & \(\mu = 0.056, \sigma = 0.053\) \\
Lognormal & \(10\) & \(1.8\) & \(\mu = -3.45, \sigma = 1.19\) \\
Gamma & \(9.9\) & \(2.3\) & \(A = 0.99, B = 0.056\) \\
Weibull & \(9.9\) & \(2.2\) & \(A = 0.056, B = 0.99\) \\
Rayleigh & \(34\) & \(40\) & \(B = 0.054\) \\
Exponential & \(9.9\) & \(2.3\) & \(\lambda = 0.056\) \\ \hline

\multicolumn{4}{c}{\textbf{28 GHz, \(\theta_b =40^\circ\)}} \\ \hline
Normal & \(18\) & \(10\) & \(\mu = 0.0523, \sigma = 0.052\) \\
Lognormal & \(8.1\) & \(1.5\) & \(\mu = -3.48, \sigma = 1.09\) \\
Gamma & \(9.5\) & \(2.5\) & \(A = 1.07, B = 0.048\) \\
Weibull & \(8.6\) & \(2.1\) & \(A = 0.052, B = 1.01\) \\
Rayleigh & \(36\) & \(44\) & \(B = 0.052\) \\
Exponential & \(8.1\) & \(1.9\) & \(\lambda = 0.052\) \\ \hline

\multicolumn{4}{c}{\textbf{28 GHz, \(\theta_b =60^\circ\)}} \\ \hline
Normal & \(21\) & \(15\) & \(\mu = 0.05, \sigma = 0.05\) \\
Lognormal & \(5.7\) & \(0.9\) & \(\mu = -3.56, \sigma = 1.09\) \\
Gamma & \(9.8\) & \(3.4\) & \(A = 1.0038, B = 0.05\) \\
Weibull & \(8.6\) & \(2.5\) & \(A = 0.049, B = 0.96\) \\
Rayleigh & \(40\) & \(62\) & \(B = 0.05\) \\ 
Exponential & \(9.8\) & \(3.3\) & \(\lambda = 0.05\) \\ \hline

\end{tabular}
\label{robot_28}
\end{table}
\subsubsection{\textcolor{black}{Unified \ac{RCS} Distribution}}
\textcolor{black}{In Fig. \ref{summary_statistical}, the best distribution is selected over all frequencies and \(\theta_b\) to formulate a histogram of all $4$ targets with their underlying \ac{RCS} distribution, based on the findings in  Tables II-XVII. We see that the lognormal distribution is the selected distribution for Mavic and \ac{QR}, whereas the gamma distribution is the selected distribution for Matrice and \ac{RA}. Interestingly, both distributions are special instances of the more general generalized Gamma distribution, whose \ac{PDF} is}
\begin{equation}
	f(x ; a, d, p)=\frac{p}{a^d \Gamma(d / p)} x^{d-1} e^{-(x / a)^p}, \quad x>0,
\end{equation} 
where $a,d,p$ denote the scale, shape, and power parameters and $\Gamma(\cdot)$ is the gamma function.
When $p = 1$, we get the gamma distribution and when $p \rightarrow 0$, we get the lognormal distribution.
\textcolor{black}{ $p$ controls the skewness of the distribution. Notice that the Mavic is a lightweight and small drone, with rapid movements and surface material of plastic, magnesium, and carbon fiber. In addition, the \ac{QR}, even though heavy, is a more dynamic target as compared to the \ac{RA}. For Mavic and \ac{QR}, lognormal appears to be the best-fit as reflections influenced by \ac{RCS} can cover several orders of magnitude, whereas for more rigid bodies such as Matrice and \ac{RA} the gamma distribution is the dominant distribution.}
\vspace{-2mm}
\section{Framework for Deterministic Radar Cross Section Evaluation}\label{sec3}
This section outlines a framework for the deterministic evaluation of the {\ac{RCS}}. 
\subsection{Measurement Setup}\label{sec3a}
%
\textcolor{black}{The \ac{InH} environment, shown in Figs. {\ref{fig1}(c)}, is an indoor corridor nearby the KINESIS lab, which is \(2.5\) m in width, \(28.5\) m in length and \(2.5\) m in height.}
 The target is a rectangular sheet \textcolor{black}{of laminated wood} with dimensions of \(1.84\) m \(\times 1.2\) m. The {\ac{NF}} distance for the target is calculated as \(D = \frac{2S^2}{\lambda}\) \cite{4154040}, where \(D\) represents the {\ac{NF}} distance, \(S\) is the largest dimension of the target (\(1.84\) m in this case), and \(\lambda\) is the operating wavelength. For the given frequency range of \(25 - 28\) GHz, the selected target's {\ac{NF}} distance lies approximately between \(564\) m and \(632\) m.

The central point of the line segment joining the {\ac{Tx}} and the {\ac{Rx}} is \((0, 0)\). The {\ac{Tx}} is positioned at \((-a, 0)\), while the {\ac{Rx}} is located at \((a, 0)\), with \(a = 0.7\) m. Target is placed at \((0, y)\), where \(y\) varies between \(2\) m and \(10\) m. The distance between the {\ac{Tx}} and the target, as well as the {\ac{Rx}} and the target, is computed as \(d = d_\mathrm{Tx, tar} = d_\mathrm{Rx, tar} = \sqrt{a^2 + y^2}\). Furthermore, the bistatic angle, which is a function of \(d\) and expressed in degrees for a target center located at \(y\) as:
\begin{equation}
    \theta_b(d) = \left(1-2\left(\frac{a}{d}\right)^2\right)\cdot\frac{180^\circ}{\pi}.
\end{equation}
It may be noted that for different \(d\) considered in the measurements, \(\theta_b(d)\) varies between approximately \(38.37^\circ -7.97^\circ\). 

The same {\ac{Tx}} and {\ac{Rx}} {\acp{RFIC}} are utilized for these measurements with an identical {\ac{IF}} value as employed in the statistical {\ac{RCS}} modeling. Additionally, identical {\ac{ZC}} sequences are transmitted. However, at the {\ac{Rx}}, {\ac{PL}} is directly obtained. All the measurements are conducted for {\ac{H}}-{\ac{H}} polarizations considering {\ac{LoS}} conditions. Moreover, the noise floor of our measurement setup is \(-72\)dB. 
 \begin{figure}[t]
    \centering
    \includegraphics[width=0.4\textwidth]{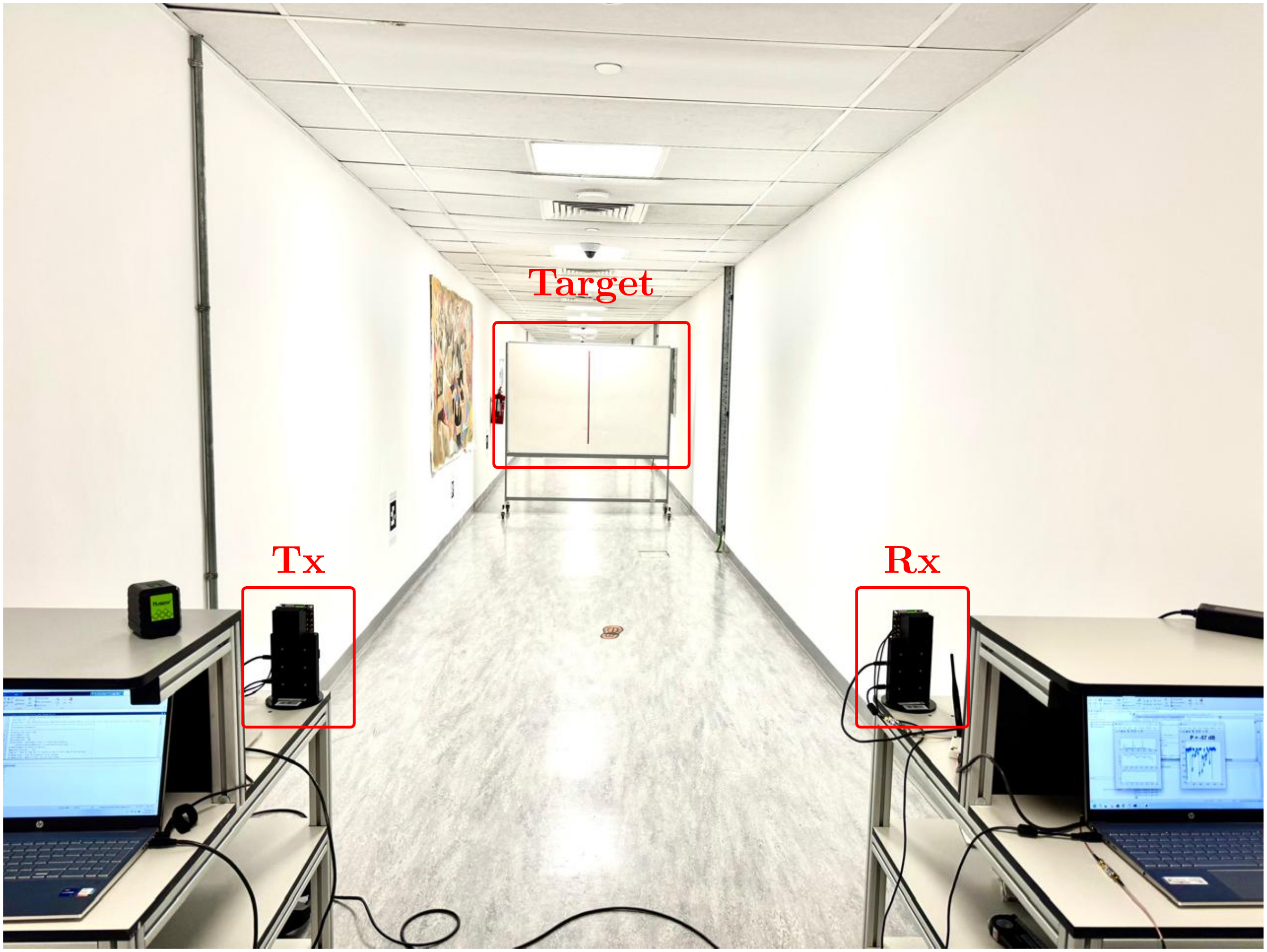} 
    \caption{The {\ac{InH}} scenario for deterministic {\ac{RCS}} evaluation.}
    \label{inh_setup} 
\end{figure}
\vspace{-0.5cm}
\subsection{{\ac{RCS}} Analysis}\label{sec3b}
\textcolor{black}{The \ac{PL} equation for the bistatic system where $d = d_\mathrm{Tx, tar} = d_\mathrm{Rx, tar}$ is}
\begin{equation}\label{pl_model}
    \mathrm{PL}_i(d,\lambda) = \alpha + 20n\log_{10}(d) -10\log_{10}(\sigma_i(d,\lambda)) + X_\sigma,
\end{equation}
where \(\alpha\) is the intercept acting as a bias, \(n\) is the {\ac{PL}} exponent, \(\sigma_i(d,\lambda)\) is the \(i\)th {\ac{RCS}} model, and \(X_\sigma\) represents the shadowing effect.
\textcolor{black}{Similar to \cite{108383}, the term in the single-way \ac{PL}, namely $10n\log_{10}(d)$, is accounted for twice due to the two-way propagation, i.e. \ac{Tx} towards target, then target towards \ac{Rx}. The intercept term is an offset that is independent of $d$ and $\lambda$ and the \ac{RCS} term depends on the material, size and shape of the target. Following \cite{4154040}, the {\ac{NF}} \ac{RCS} of the sheet depends on both distance and wavelength, because, in the \ac{NF} region, the amount of energy reflected back towards the \ac{Rx} varies with distance, as the signal may not fully illuminate the target at different \ac{NF} distances. We explore multiple \ac{NF} \ac{RCS} models for the rectangular sheet, as}
%
\begin{equation}\label{approx_rcs}
\begin{split}
\sigma_1(d,\lambda) &= a_1d^2\cos^m(\theta_b(d)),\\
\sigma_2(d,\lambda) &= \left(a_1d^2+a_2\lambda d^3\right)\cos^m(\theta_b(d)),\\
\sigma_3(d,\lambda) &= \left(a_1d^2+a_2\lambda d^3+a_3\lambda^2 d^4\right)\cos^m(\theta_b(d)),
\end{split}
\end{equation}
where \( \sigma_1(d, \lambda) \), \( \sigma_2(d, \lambda) \), and \( \sigma_3(d, \lambda) \) are the proposed bistatic {\ac{RCS}} models for the rectangular sheet as functions of the distance \(d\) and the wavelength \(\lambda\). \( a_1, a_2, \) and \( a_3 \) are fitting parameters that characterize the {\ac{RCS}} based on the geometric properties of the target. 

\begin{figure}[t]
    \centering
    \includegraphics[trim={5mm 5mm 5mm 6mm},clip,width=0.37\textwidth]{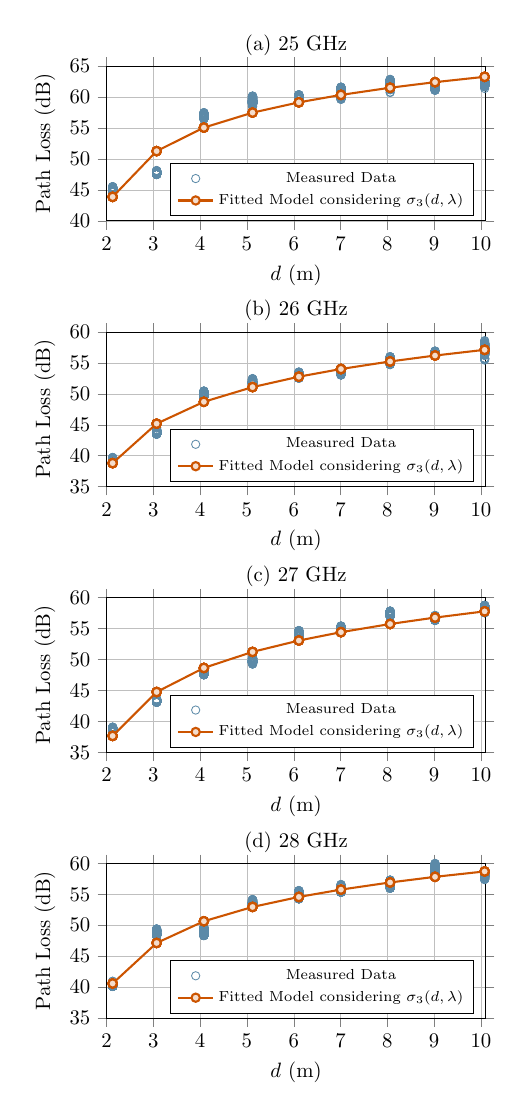} 
    \caption{Measured {\ac{PL}} and model curve fitting in the {\ac{NF}}. }
    \label{result_deter_rcs} 
\end{figure}

\textcolor{black}{To justify equation \eqref{pl_model}, we plot in Fig. \ref{result_deter_rcs} (a)-(d) the {\ac{PL}} as function of distance, which as observed does not follow a linear growth pattern but instead reaches a plateau as \(d\) increases. This PL plateau is a characteristic of the \ac{NF} regime, arising from the non-linear behavior of the \ac{NF} \ac{RCS} of the target. This trend remains consistent across the \(25 - 28\) GHz frequency range. Additionally, the fitted model corresponding to \(\sigma_3(d,\lambda)\) demonstrates the lowest mean fitting error \ac{MFE} among all the considered RCS models (see Table \ref{rcs_dterm}).}

Since we have used only one rectangular sheet for our measurement, we cannot characterize the dependence of \( a_1, a_2, \) and \( a_3 \) on the area of the sheet. \( \cos^m(\theta_b(d)) \) captures the angular dependence of the bistatic {\ac{RCS}}, where \(\theta_b(d)\) is the bistatic angle as a function of distance \(d\), and \(m\) is an angular exponent. \( d^2, \lambda d^3, \) and \( \lambda^2 d^4 \) represent the terms that account for the {\ac{NF}} scattering behavior of the rectangular sheet at different orders of distance and wavelength. 

To evaluate the {\ac{GoF}}, {\ac{MFE}} is evaluated for each {\ac{RCS}} model, where {\ac{MFE}} in percentage is defined as
\begin{equation}
    \mathrm{MFE} = \frac{1}{N} \sum\nolimits_{j=1}^N \left| \frac{\mathrm{PL}_{\mathrm{meas},j} - \mathrm{PL}_{i,j}}{\mathrm{PL}_{\mathrm{meas},j}} \right| \times 100\%
\end{equation}
where \(N\) is the total number of data points, \(\mathrm{PL}_{\mathrm{meas},j}\) is the measured {\ac{PL}} for the \(j\)-th data point, \(\mathrm{PL}_{i,j}\) is the \(i\)th modeled {\ac{PL}} for the \(j\)-th data point. The primary distinction among the various {\ac{PL}} models lies in the selection of the {\ac{RCS}} model employed. Table \ref{rcs_dterm} presents the results of the fitting process for the different {\ac{PL}} models across various frequencies. \(X_\sigma\)is evaluated as \(\sqrt{\frac{1}{N} \sum\nolimits_{j=1}^N (x_j - \mu)^2}\), where \(x_j\) represents the difference between the measured values and the fitted curve values, and \(\mu\) denotes the average value of \(x_j\). Furthermore, the fitted models based on the measured data are depicted in Fig. \ref{result_deter_rcs} (a)-(d) for the operating frequencies ranging from \(25-28\) GHz. To facilitate interpretation, the {\ac{PL}} values are inverted by multiplying them by \(-1\) in Fig. \ref{result_deter_rcs} (a)-(d), effectively representing the negative of the {\ac{PL}}. 

\begin{table}[t]
\centering
\small
\caption{Fitting parameters for deterministic {\ac{RCS}}.}
\begin{tabular}{p{0.9cm}|p{0.6cm}|p{0.4cm}|p{0.7cm}|p{0.4cm}|p{0.4cm}|p{0.4cm}|p{0.4cm}|p{0.7cm}}
\hline
\multicolumn{9}{c}{\textbf{25 GHz}} \\ \hline
\textbf{Approx.} & \(\alpha\) & \(n\) & \(m\) & \(a_1\) & \(a_2\) & \(a_3\) & \(X_\sigma\) & \textbf{MFE} \\ \hline
\(\sigma_1(d,\lambda)\)&\(51.41\)&\(1.85\)&\(-7.86\)&\(2.96\)&\(-\)&\(-\)&\(1.64\)&\(2.400\)\\ \hline
\(\sigma_2(d,\lambda)\)&\(51.66\)&\(1.84\)&\(-8.13\)&\(2.91\)&\(1.25\)&\(-\)&\(1.63\)&\(2.377\) \\ \hline
\(\sigma_3(d,\lambda)\)&\(51.82\)&\(1.83\)&\(-8.26\)&\(2.90\)&\(1.25\)&\(0.03\)&\(1.63\)& \(2.370\)\\
\hline
 \multicolumn{9}{c}{\textbf{26 GHz}} \\ \hline
 \(\sigma_1(d,\lambda)\)&\(44.21\)&\(1.90\)&\(-6.1\)&\(2.94\)&\(-\)&\(-\)&\(0.72\)&\(1.146\)\\ \hline
\(\sigma_2(d,\lambda)\)&\(43.80\)&\(1.92\)&\(-6.02\)&\(2.90\)&\(1.25\)&\(-\)&\(0.72\)& \(1.144\)\\ \hline
\(\sigma_3(d,\lambda)\)&\(43.65\)&\(1.92\)&\(-6.02\)&\(2.80\)&\(1.25\)&\(0.03\)&\(0.72\)&\(1.144\) \\ \hline
 \multicolumn{9}{c}{\textbf{27 GHz}} \\ \hline
 \(\sigma_1(d,\lambda)\)&\(43.78\)&\(1.95\)&\(-7.19\)&\(2.99\)&\(-\)&\(-\)&\(2.09\)&\(1.813\)\\ \hline
\(\sigma_2(d,\lambda)\)&\(44.17\)&\(1.97\)&\(-7.04\)&\(3.50\)&\(1.01\)&\(-\)&\(2.03\)&\(1.791\) \\ \hline
\(\sigma_3(d,\lambda)\)&\(43.99\)&\(1.98\)&\(-6.91\)&\(3.50\)&\(1.01\)&\(0.02\)&\(2.02\)&\(1.767\) \\ 
 \hline
 \multicolumn{9}{c}{\textbf{28 GHz}} \\ \hline
 \(\sigma_1(d,\lambda)\)&\(46.80\)&\(1.85\)&\(-6.56\)&\(3.02\)&\(-\)&\(-\)&\(1.04\)&\(1.597\)\\ \hline
\(\sigma_2(d,\lambda)\)&\(47.73\)&\(1.85\)&\(-6.68\)&\(3.25\)&\(1.01\)&\(-\)&\(1.04\)& \(1.582\)\\ \hline
\(\sigma_3(d,\lambda)\)&\(47.56\)&\(1.85\)&\(-6.69\)&\(3.50\)& \(1.01\)&\(0.02\)&\(1.04\)& \(1.583\)\\ 
\hline
\end{tabular}
\label{rcs_dterm}
\end{table}
Table \ref{rcs_dterm} presents the fitting parameters for deterministic {\ac{RCS}} models \((\sigma_1(d,\lambda), \sigma_2(d,\lambda)\), and \(\sigma_3(d,\lambda)\) across operating frequencies of \(25\) GHz, \(26\) GHz, \(27\) GHz, and \(28\) GHz. The intercept, \(\alpha\) and {\ac{PL}} exponent, \(n\), show minor variations across frequencies and models, with \(\alpha\) decreasing slightly as frequency increases (e.g., from \(51.41\) at \(25\) GHz to \(47.56\) at \(28\) GHz) and \(n\) remains consistent, ranging from \(1.83\) to \(1.98\). The angular exponent, \(m\), varies across frequencies, ranging from \(-8.26\) at \(25\) GHz for \(\sigma_3(d,\lambda)\) to \(-6.69\) at \(28\) GHz for \(\sigma_3(d,\lambda)\), indicating its sensitivity to frequency. The {\ac{RCS}} model parameters, \(a_1, a_2\), and \(a_3\) are consistent within each frequency and model. The small values of the parameters \(a_2\) and \(a_3\) indicate their minor contributions towards the deterministic {\ac{RCS}}. However, it is again highlighted that \(a_2\) and \(a_3\) also depend on the area of the rectangular sheet. The change in the area of the sheet would certainly change the values of \(a_2\) and \(a_3\). \textcolor{black}{The} \(a_2\) \textcolor{black}{and} \(a_3\) parameters are expected to be higher for rectangular sheets with smaller areas. Among the models, \(\sigma_3(d,\lambda)\) generally provides the lowest {\ac{MFE}}, demonstrating better fitting performance. The {\ac{MFE}} values decrease with increasing frequency, indicating improved fitting accuracy at higher frequencies. Overall, \(\sigma_3(d,\lambda)\) is the most accurate model across all frequencies due to its inclusion of higher-order terms and consistently lower {\ac{MFE}} values. 
\vspace{-0.25cm}
\section{Conclusions}\label{sec4}
This study comprehensively characterizes the {\ac{RCS}} of typical targets in {\ac{InF}} and {\ac{InH}} environments using statistical and deterministic modeling approaches. The statistical analysis involved fitting measured {\ac{RCS}} data from test targets such as {\acp{UAV}}, {\ac{RA}}, and {\ac{QR}} to parametric distributions. The statistical analysis is performed for monostatic configuration as well as three different bistatic configurations for operating frequencies of \(25 - 28\) GHz. Weibull, lognormal, and gamma distributions, all of which are right-skewed distributions, were identified as optimal based on {\ac{GoF}} metrics, effectively capturing the statistical variations in {\ac{RCS}} induced by target geometry, material composition, dynamic operational conditions, etc. Furthermore, it has been observed that the {\ac{RCS}} decreases with an increase in the bistatic angle. In the deterministic analysis, the bistatic {\ac{RCS}} of a rectangular sheet \textcolor{black}{of laminated wood} was modeled by leveraging {\ac{PL}} measurements across frequencies from \(25\) GHz to \(28\) GHz under {\ac{NF}} conditions. Proposed {\ac{RCS}} models, parameterized by bistatic angle and the distance between the {\ac{Tx}}, {\ac{Rx}}, and the target, were evaluated, and the best-fitting models were identified based on the {\ac{MFE}}. The results indicate that {\ac{RCS}} strongly depends on the bistatic angle and target distance, emphasizing the dependence of angular and spatial variability inherent in bistatic scattering on the {\ac{RCS}}, \textcolor{black}{which further} highlights the critical role of {\ac{RCS}} modeling in channel modeling for \ac{ISAC}. Future work may focus on extending these models to other frequencies and environments.
\vspace{-0.5cm}
\bibliographystyle{unsrt}
\bibliography{main}

\end{document}